\documentclass[seceq]{ptptex}

\usepackage{graphicx}

\title{
An Alternative Understanding of Mass Formulas \\
 in Terms of Nuclear Structure
}

\author{
Munetake \textsc{Hasegawa}$^1$
and Kazunari \textsc{Kaneko}$^2$
}

\inst{
$^1$Laboratory of Physics, Fukuoka Dental College,
 Fukuoka 814-0193 \\
$^2$Department of Physics, Kyushu Sangyo University,
 Fukuoka 813-8503 \\
}

\abst{

   A typical form of mass formula is re-explained in terms of nuclear structure.
 For $ N \approx Z $ nuclei, we propose to start with the shell model
 picture and to consider the $T=0$ $2n-2p$ ($\alpha$-like) correlations
 as the fundamental concept, instead of the symmetry energy.
 Subsequently, the symmetry energy is described on the basis of
 the $\alpha$-like superfluidity caused by the $T=0$ $2n-2p$ correlations,
 in parallel with the pairing energy
  described on the basis of the pairing superfluidity.
 This re-explanation gives useful insight for understanding the nuclear
 mass formula. The origin of the Wigner energy is also explained
 in an interacting boson model for the Cooper pairs in the $\alpha$-like
 superfluid vacuum.
 Adding a correction term due to the $T=0$ $2n-2p$ correlations,
 which determines the $T=0$ base level for nuclear masses,
 can improve the mass formulas in practice.

}

\begin{document}

\maketitle

\section{Introduction}

   The Weizs\"acker-Bethe mass formula has played an important role
 for many years.
 It offers a guideline for developing modern mass formulas.
 Indeed, modern mass formulas, such as the droplet mass formula (FRDM)
  \cite{Moller}
 (excluding the Hartree-Fock approaches \cite{Aboussir,Goriely})
 largely retain the original form.
  They include the pairing and symmetry energy terms, in addition to
 the volume, surface and Coulomb energy terms. While the pairing energy
 has been investigated by combining with the microscopic pairing correlations,
 the symmetry energy has not necessarily been investigated from the point
 of view of the shell model.
   A recent detailed work \cite{Duflo} succeeded in obtaining a precise
 mass formula.  That work is based on the rigorous microscopic guidelines
 given in Ref. 5), which considers the monopole field and pairing
 structure providing the dominant terms of the mass formula.
 The study presented in Ref. 5) considers general properties of
 the shell model Hamiltonian. However, while the pairing interaction
 as the origin of the pairing energy determines the structure of wave functions,
 the symmetry energy is not treated symmetrically with the pairing energy.
 
   In the present paradigm for the nuclear mass formula, the symmetry energy
 is regarded as a basic concept. In the shell model, however,
 there is no approach other than describing the symmetry energy in terms of
 nuclear correlations. The asymmetrical treatment of the symmetry energy
 and pairing energy leaves a missing link in relating mass formulas
 to nuclear structure.
  This paper proposes an alternative approach, in which the ``symmetry energy"
 is not treated as a fundamental concept and explains the symmetry energy
  in terms of certain correlations, as the pairing energy is determined 
 in terms of the pair correlations.  We wish to understand the symmetry energy
 derived from the mean field theory from the point of view of the shell model.
 Our treatment begins from the $jj$ coupling shell model
 based on a $Z=N$ doubly-closed shell core, and we do not discuss Strutinski's
 prescription \cite{Strut}.
  The purpose of this paper is not to give a new mass formula better than
 modern sophisticated mass formulas, but to present a useful understanding
 of the mass formula.  We therefore start from an old fashioned mass formula
 in order to clearly show the basic idea.

  In the $jj$ coupling shell model with an effective interaction,
 the energy depending on the total isospin $T$ comes from the interactions
 between valence nucleons in $j$ orbits. The corresponding correlations
 are not yet reduced to a mean field but determine wave functions
 or structure of nuclei, in the shell model description.
 We consider such correlations in even-even $N=Z$ ($A_0+m\alpha$) nuclei
 that give no contribution of the symmetry energy.
 Here, $A_0$ represents a doubly-closed-shell core, $\alpha$ is
 a two-neutron-two-proton ($2n-2p$) quartet with $T=0$, and $m$ is an integer.
 We show that the interaction energies of the $A_0+m\alpha$
 nuclei characterize the binding energies of $N \approx Z$ nuclei,
 excluding the bulk energy depending on mass $A$.
 The strength of the correlations can be evaluated in terms of the difference
 between the mass of an $A_0+m\alpha +2n$ ($A_0+m\alpha +2p$) nucleus
 and the average mass
 of its neighboring nuclei with $A=A_0+m\alpha$ and $A=A_0+(m+1)\alpha$.
 With this indicator, Gambhir, Ring and Schuck \cite{Gamb} studied
 a superfluid state of many $\alpha$ {\it particles}.
 The term $\alpha$, however, represents only a $T=0$ $2n-2p$ quartet,
 not the spatial $\alpha$ cluster. We call the correlations
 ``$T=0$ $2n-2p$ correlations" in the sense of many-body correlations.
 (We use ``$\alpha$-like" as a concise term for the superfluid state.)
 This paper shows that the symmetry energy is derived from the nonparticipation
 of redundant nucleons in the $T=0$ $2n-2p$ correlations,
 in parallel with the pairing energy derived from the nonparticipation
 of an odd nucleon in the $T=1$ pair correlations.

  The energy of valence nucleons is separated from the binding energy,
  and the leading role of $T=0$ $2n-2p$ correlations is discussed in $\S$2.
 Section 3 discusses the fundamental $T=0$ $2n-2p$ correlated structure
 in $N \approx Z$ nuclei, (which is called ``$\alpha$-like superfluidity").
 Section 4 explains the mass differences between even-even nuclei
 in terms of multi-pair structure on the base of $\alpha$-like superfluidity.
 In $\S$5, we discuss  how the pairing energy should be evaluated.
 Section 6 gives concluding remarks.

\section{Correlations of valence nucleons buried in the binding energy}

\subsection{Extraction of the energy of valence nucleons}

   In the old fashioned mass formula, the bulk of the binding energy is
 written in terms of the volume, surface and Coulomb energies as
\begin{equation}
  B_{VSC}(A) = - a_V A + a_S A^{2/3} + a_C Z^2 / A^{1/3}.  \label{eq:1}
\end{equation}
 We can consider that these main terms basically represent a nuclear
 potential in the shell model picture, while the other terms of the mass
 formula are related to the shell model interactions.
 It must be stressed that the symmetry energy depending on the total
 isospin $T$ is attributed to the shell model interactions in this picture.
 Let us estimate the interaction energy by subtracting $B_{VSC}(A)$
 from the experimental binding energy $B(A)$ \cite{Audi} for $A_0+m\alpha$
 nuclei with $T=0$.  (Note that the sign of the binding energy $B(A)$ is
 negative in this paper.)
 The values $B(A)-B_{VSC}(A)$ calculated with a few mass formulas with simple
 forms \cite{Duflo,Yagi,Ring,Samanta} are listed in Table \ref{table1}.
 [In the third line, we used the six-parameter mass formula in Ref. 4).
  The Coulomb energy term in Refs. 4) and 11) is expressed in terms of
 different functions of the proton number $Z$.]
 In Table \ref{table1}, we tabulate the values $B(A)-B_{VSC}(A)$
 for seven $A=A_0+m\alpha$
 nuclei with $T=0$, where the symmetry energy makes no contribution.
 These values display variation depending on $A$.
 
\begin{table}[b]
\caption{The values of $B(A)-B_{VSC}(A)$ for $A_0+m\alpha$ nuclei with $T=0$,
         calculated using a few mass formulas.}
\begin{center}
\begin{tabular}{c|rrrrrrr}   \hline
  ref. & $^{16}$O & $^{20}$Ne & $^{28}$Si & $^{40}$Ca & $^{44}$Ti & $^{56}$Ni
       & $^{64}$Ge \\ \hline
\cite{Yagi}    &  $-5.81$ &  $-2.13$ &  $-4.82$ &  $-2.88$ &  $-1.35$ &  $-7.99$ &    $-4.63$  \\
\cite{Ring}    & $-12.80$ & $-10.21$ & $-14.92$ & $-15.79$ & $-15.17$ & $-24.50$ &    $-22.94$  \\
\cite{Duflo}   &  $-4.85$ &  $-0.87$ &  $-2.97$ &  $-0.15$ &  $ 1.67$ &  $-4.10$ &    $-0.17$  \\
\cite{Samanta} &  $-7.21$ &  $-3.52$ &  $-6.08$ &  $-3.85$ &  $-2.21$ &  $-8.49$ &    $-4.91$  \\
   \hline
\end{tabular}
\end{center}
\label{table1}
\end{table}

   Table \ref{table1} shows that the mass formula \cite{Ring} fitted
 for heavy nuclei is not good for $N \approx Z$ nuclei but that the other three
 display parallel and interesting behavior (dips at $^{28}$Si and
 $^{56}$Ni) as $A$ increases.
 According to ordinary mass formulas, there remains the pairing term $\delta_P$,
 which contributes to the $T=0$ even-even nuclei under consideration.
 However, the deviations of the experimental binding energies from $B_{VSC}$
 shown in Table \ref{table1} are much larger than the pairing effect.
 The characteristic behavior of $B(A_0+m\alpha)-B_{VSC}(A_0+m\alpha)$
 cannot be explained as a simple variation of $\delta_P$ depending on $A$,
 like $\delta_P \propto A^p$.  The existing mass formulas do not describe
 the behavior.  The characteristic behavior of
  $B(A_0+m\alpha)-B_{VSC}(A_0+m\alpha)$
 must reflect correlations stronger than the pairing correlations
 from the point of view of the shell model.
 This is worth investigating further.
  Let us start from the old simple mass formula \cite{Yagi}
 with the parameters $a_V=15.56$, $a_S=17.23$ and $a_C=0.6986$ in MeV.
  It is noticed in Table \ref{table1} that the values
 in the first line \cite{Yagi} resemble those in the fourth line obtained
 with the mass formula \cite{Samanta}, and the mass formula \cite{Duflo}
 has an elaborate form, so that the deviations
  $B(A_0+m\alpha)-B_{VSC}(A_0+m\alpha)$ may be small.
 
\begin{figure}[t]
\begin{center}
\includegraphics[width=7cm,height=7cm]{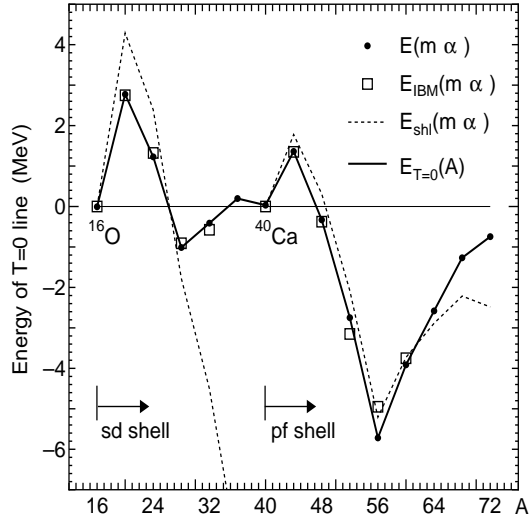}
  \caption{The energies of $E(A_0+m\alpha)=B(A_0+m\alpha)-B_0(A_0+m\alpha)$
           and the curve connecting them, which defines the $T=0$ base level.}
  \label{fig1}
\end{center}
\end{figure}

    The values in Table \ref{table1} indicate the insufficiency of $B_{VSC}$
 for the doubly-closed-shell nuclei $^{16}$O and $^{40}$Ca, which are
 the bases for the shell model calculations.  We suppose that the deviations
 in $^{16}$O and $^{40}$Ca require adjustments of the depth of the shell model
 potential.  The adjustment parameter $\delta_P$ for even-even nuclei has
 the form $A^{-3/4}$ in the old convention.  If we adopt the $A^{-3/4}$
 adjustment for the potential depth, we can fix its parameter
 so as to make the deviations $B(A_0+m\alpha)-B_{VSC}(A_0+m\alpha)$
 nearly zero for $^{16}$O and $^{40}$Ca.
 We assume that the main part of the mass formula corresponding to
 the shell model potential is approximated by
\begin{eqnarray}
  & {} &          B_0(A) = B_{VSC}(A) + \delta U_{pot}(A), \\ \label{eq:2}
  & {} & \delta U_{pot}(A) = - 46.4 / A^{3/4}.   \label{eq:3}
\end{eqnarray}
 The deviation $B(A)-B_0(A)$ could be regarded as the energy of valence
 nucleons outside a doubly-closed-shell core,
\begin{equation}
  E(A) = B(A) - B_0(A).  \label{eq:4}
\end{equation}
 The energies $E(A_0+m\alpha)$ for the even-even $N=Z$ nuclei with $T=0$
 are plotted in Fig. \ref{fig1}, which displays the characteristic behavior
 of the binding energies $B(A_0+m\alpha)$ mentioned above.
 
   In the shell model calculation, the experimental energy of correlated
 valence nucleons outside a doubly-closed-shell core $A_0=(N_0,Z_0)$ is
 evaluated using
\begin{equation}
  E_{shl}(N,Z) = B(N,Z) - B(N_0,Z_0)
               - \lambda (A-A_0) - \Delta E_C(N,Z),  \label{eq:5}
\end{equation}
 where $\Delta E_C(N,Z)$ is a Coulomb energy correction for the valence
 nucleons.  For instance, the correction
  $\Delta E_C(N,Z) = p(Z-Z_0)+q(Z-Z_0)(Z-Z_0-1)+r(Z-Z_0)(N-N_0)$ 
 was used by Caurier et al. \cite{Caurier} in the shell model calculations
 for $f_{7/2}$ shell nuclei.  We calculated the energy $E_{shl}(N,Z)$
 using the parameter values $p=7.279$, $q=0.15$, $r=-0.065$ and $\lambda =-12.45$
 (in MeV) for the $pf$ shell nuclei and $p=3.54$, $q=0.20$, $r=0.0$ and
 $\lambda =-11.2$ (in MeV) for the $sd$ shell nuclei.  The calculated values
 are indicated by the dotted curves in Fig. \ref{fig1}.  We can see that
 $E(A_0+m\alpha) \approx E_{shl}(A_0+m\alpha)$ in the first half of the shells,
 which supports our assumption that $E(A)$ in Eq. (\ref{eq:4}) represents
 the energy of valence nucleons. 
 The disagreement between $E(A_0+m\alpha)$ and $E_{shl}(A_0+m\alpha)$ is large
 in the latter half of the shells. (Note that the disagreement is smaller
 in the heavier $pf$ shell nuclei.)  The hole picture
 would be better for the latter half of the shells.

\subsection{The leading role of the $T=0$ $2n-2p$ correlations}

\begin{figure}[t]
\begin{center}
\includegraphics[width=7.5cm,height=7.5cm]{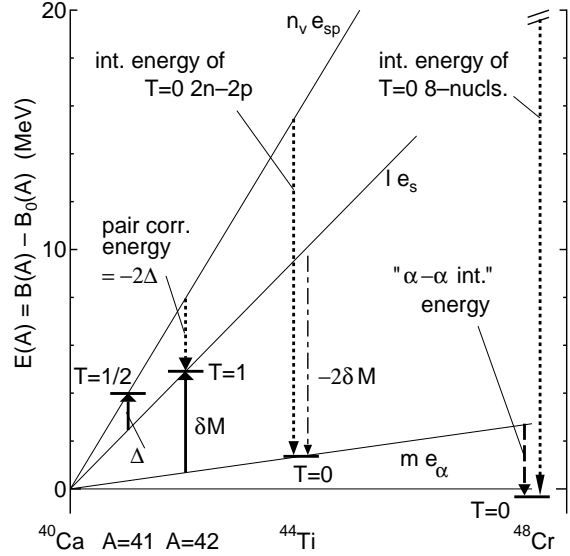}
  \caption{Schematic depiction of the pair correlations and
           $T=0$ $2n-2p$ correlations and their indicators
           for the nuclei $^{40}$Ca, $A=41$ with $T=1/2$, $A=42$ with
           $T=1$, $^{44}$Ti and $^{48}$Cr.}
  \label{fig2}
\end{center}
\end{figure}

   From the above, we find that the energy $E(A)$ defined by Eq. (\ref{eq:4})
 approximately represents the total energy of valence nucleons outside
 the doubly-closed-shell core $^{16}$O or $^{40}$Ca.  To obtain a guide
 for the discussion given in the following sections, let us consider
 the main features of $E(A)$ near $^{40}$Ca,
  where $E(A)$ corresponds well with the shell model
 energy $E_{shl}(A)$.  Figure \ref{fig2} depicts the ground state energies 
 $E(A)$ of $^{40}$Ca, an $A=41$ system with $T=1/2$, an $A=42$ system with $T=1$,
 $^{44}$Ti and $^{48}$Cr. 
 (This figure is similar to the diagram for the pairing vibrations.) 
  The energy $E(A)$ in Eq. (\ref{eq:4}) does not
 give exactly the same energy to $^{41}$Ca and $^{41}$Sc.  A correction term
 representing the Coulomb energy is necessary in the final stage.
  In Fig. \ref{fig2},
 we show the average energy of $E(^{41}{\rm Ca})$ and $E(^{41}{\rm Sc})$
 for the $A=41$ system with $T=1/2$.  Similarly, we show the average energy
 of $E(^{42}{\rm Ca})$ and $E(^{42}{\rm Ti})$
 [which is approximately equal to $E(^{42}{\rm Sc})$]
  for the $A=42$ system with $T=1$.
   The energy $E(A=41)$ represents
 an effective single-particle energy $e_{sp}$ in the nuclear potential
 represented by $B_0(A)$, whose depth is adjusted to be zero for
 $^{40}$Ca.  If there are no interactions between valence nucleons,
 the energy of the $A=40+n_v$ system ($n_v$ being the number of valence nucleons)
 is $n_v e_{sp}$. However, the real energy $E(A=42)$ lies substantially below
 the line $n_v e_{sp}$ in Fig. \ref{fig2}.  The difference is the pair correlation
 energy. Half of the absolute pair correlation energy is called the
 (three-point) odd-even mass difference $\Delta$.  The value $\Delta$ is
 often used as the indicator of the pair correlations.  We show the odd-even mass
 difference $\Delta$ at $A=41$ in Fig. \ref{fig2}.  The definition of $\Delta$
 given in Eq. (\ref{eq:30}) explains the geometrical relations
  shown in Fig. \ref{fig2}.

   The energy of $^{44}$Ti measured from the line $n_v e_{sp}$ is the interaction
 energy of the four nucleons with $T=0$ outside the $^{40}$Ca core,
 which is denoted by the dotted line in Fig.~\ref{fig2}.  Figure \ref{fig2} shows
 that the $T=0$ four nucleon correlations, which we call $T=0$ $2n-2p$ correlations,
 experience an energy gain much larger than
  that of the pair correlations $2 \Delta$.
 The strong $T=0$ $2n-2p$ correlations
 cause the four nucleons to form an $\alpha$-like quartet outside the $^{40}$Ca
 core, as described by the core plus $\alpha$-cluster model.  The study of
 $\alpha$-like correlations has a long history.
 \cite{Maru,Danos,Arim,Kamim,Tomo,Cauvin,Duss1,Gamb,Jensen,Apos,Duss2,Curut,Suga,Hase1,Hase1B,Merm,Hase2,Sakuda}
  (Also, see other references cited in Ref. 28), especially for the core
  plus $\alpha$-cluster model.)
   Sometimes the indicator of the $T=0$ $2n-2p$ correlations was evaluated
  after subtracting the symmetry energy from the binding energy
 (for instance, see Ref. 20)).  As shown below, however,
 different involvements of valence nucleons in
 the $T=0$ $2n-2p$ correlations result in the ordering
 of the ground-state energies of $N \approx Z$ nuclei according to that of
 the total isospin $T$ (which results in the symmetry energy
 in the mean field theory).
  If we limit the comparative study of different $T$ nuclei
 by excluding the symmetry energy from the binding energy at the beginning,
 we miss a substantial energy gain due to the underlying $T=0$ $2n-2p$
 correlations, and we therefore cannot understand the formation
 of the $\alpha$-cluster in $^{20}$Ne and $^{44}$Ti
 in contrast to 
 the absence of the $\alpha$-cluster outside the core in $^{20}$O and $^{44}$Ca.
 
   Let $e_s$ denote the energy of the $T=1$ correlated pair and $l$ the number
 of correlated pairs.  If there is no interaction between the $2n$ and $2p$
 pairs, the energy of $^{44}$Ti is expected to be on the line $l e_s$.
  The real energy of $^{44}$Ti lies far below the line $l e_s$
 in Fig. \ref{fig2}.  The difference is the interaction energy between the $2n$
 and $2p$ pairs, which is denoted by the dot-dashed line in Fig. \ref{fig2}.
 Half of the absolute interaction energy between the $2n$ and $2p$ pairs
 is a good indicator of the $T=0$ $2n-2p$ correlations.  This indicator
 is called  the ``ODD-EVEN mass difference for the $\alpha$-like correlations" 
 by Gambhir, Ring and Schuck \cite{Gamb}, where ODD and EVEN
 are used for the number of pairs.  We define it in Eq. (\ref{eq:13})
 and Eq. (\ref{eq:14}), and we write it as $\delta M(A_0+m\alpha +2)$ and
 $\delta W(A_0+m\alpha +2:T=1)$ in the respective cases.
   The value $\delta M$ is indicated by the solid line
 at $A=42$ in Fig. \ref{fig2}. 
 The definitions given in Eq. (\ref{eq:13}) and Eq. (\ref{eq:14}) explain
 the geometrical relations.  The total interaction energy of
 the $T=0$ $2n-2p$ quartet is $-2(2 \Delta + \delta M)$.
 Figure \ref{fig2} clearly shows that the ODD-EVEN mass difference $\delta M$
 for the $T=0$ $2n-2p$ correlations is much larger than the odd-even
 mass difference $\Delta$ for the pair correlations.  The ODD-EVEN mass
 difference $\delta M$ for the $A=42$ system with $T=1$, which is measured from
 the $T=0$ line, reflects the symmetry energy.  In other words,
 the symmetry energy
 in the framework of the mass formula can be explained in terms of
 the $T=0$ $2n-2p$ correlations from the point of view of the shell model.
 It should be noticed that the symmetry energy and pairing energy in the mass
 formula are treated on the same footing here.
 
   Because the $T=0$ $2n-2p$ correlations are very strong, we believe
 that the $T=0$ $2n-2p$ quartet is approximately a good excitation mode.
 Let its energy be $e_\alpha$. Then the energy of the $A=40+m \alpha$ system
 is expected to be nearly $m e_\alpha$.  This expectation holds roughly
 for $^{48}$Cr, as shown in Fig. \ref{fig2}, where $E(^{48}{\rm Cr})$ lies
 below but near the line $m e_\alpha$. 
 It is shown in Ref. 29) that the $^{48}$Cr nucleus is described
 quite well by the $^{40}$Ca core plus two $\alpha$-cluster model.
  Figure \ref{fig1} shows that the energies $E(40+m \alpha)$
 are below the line $m e_\alpha$.
 This result indicates the important point that the interaction between
 the $T=0$ $2n-2p$ quartets is attractive, and the $T=0$ $2n-2p$ 
 correlations are collective in systems of many quartets.
 
   Figure \ref{fig1} shows that $^{56}$Ni in the middle of the $pf$ shell
 is different from the typical doubly-closed-shell nuclei $^{16}$O and
 $^{40}$Ca, but it resembles $^{28}$Si in the middle of the $sd$ shell.
  Because $B_0(A_0+m\alpha)$ as a function of $m$ is monotonic
  in the regions $A=16-36$ and $A=40-72$ of the $A_0+m\alpha$ nuclei,
 we cannot attribute the difference between the $^{56}$Ni nucleus
 and the $^{16}$O and $^{40}$Ca nuclei to special behavior of $B_0(A)$.
  The difference is due to correlations of valence nucleons or a shell effect.
 The rigid core of $^{16}$O and $^{40}$Ca is supported by the successful
 description of $^{20}$Ne and $^{44}$Ti with the core plus $\alpha$-cluster
 model. Figure \ref{fig1} suggests structure of $^{56}$Ni ($^{28}$Si)
 that differs from a rigid core.
 We suppose that $^{16}$O and $^{40}$Ca have rather rigid cores and
 that the other $N \approx Z$ nuclei are described as systems of
 correlated valence nucleons outside the respective cores,
 as is done in ordinary shell model calculations.

\subsection{Examination by means of the shell model calculation}

   Let us examine the above picture by carrying out shell model calculations
 with a realistic effective interaction in the $pf$ shell nuclei
 outside the $^{40}$Ca core.
 The shell model Hamiltonian describing valence nucleons outside the core
 is composed of the single-particle energy part and the effective interaction:
\begin{equation}
  H = H_{sp} + H_{int}.     \label{eq:H1}
\end{equation}
 We adopt the Honma interaction, \cite{Honma} which accurately describes
  the $pf$ shell nuclei near $^{56}$Ni,
 and we consider systems in the $jj$ coupling scheme.
 To compare with Fig. \ref{fig1}, we use the same parameter value
  $\lambda = -12.45$ MeV, as in Eq. (\ref{eq:5}),
   though a somewhat different Coulomb energy correction
 is used in Ref. 30).  The adopted single-particle energies are
 $e(f_{7/2})=3.862$, $e(p_{3/2})=6.7707$, $e(p_{1/2})=8.313$ and
 $e(f_{5/2})=11.0671$ in MeV. 
 
   It is useful to decompose the effective interaction $H_{int}$ into
 the monopole part and the residual part \cite{Dufour}.  The $T=0$ monopole
 field defined by the following equation is especially important, because it
 determines the main part of the interaction energy (expectation value
 $\langle H_{int} \rangle$):
\begin{eqnarray}
 & {} & H_{mp}^{T=0} = - k^0 \sum_{a \leq b} \sum_{JM}
         A^\dagger_{JM00}(ab) A_{JM00}(ab),               \\ \label{eq:H2}
 & {} &  k^0 = \frac{\sum_{ab} \overline{V}(ab:T=0)} {\sum_{ab}1}  \label{eq:H3}
\end{eqnarray}
 with
\begin{equation}
   \overline{V}(ab:T=0) = \frac{ \sum_J (2J+1) V(abab:J,T=0) }{\sum_J (2J+1)},
                                  \label{eq:H4}
\end{equation}
 where $A^\dagger_{JMTK}(ab)$ is the creation operator of a nucleon pair
 with spin $JM$ and isospin $TK$ in the single-particle orbits
 ($a,b$) and $V(abab:JT)$ is a diagonal two-body interaction matrix element.
 Let us write the effective interaction as
\begin{equation}
  H_{int} = H_{mp}^{T=0} + H_{res} \quad ( H_{res} = H_{int} - H_{mp}^{T=0}).
        \label{eq:H5}
\end{equation}
 The monopole field $H_{mp}^{T=0}$ is expressed exactly as
\begin{equation}
 H_{mp}^{T=0} = - \frac{k^0}{2} \Big\{ \frac{\hat n_v}{2}
        \big(\frac{\hat n_v}{2} +1\big)
   - {\hat T}({\hat T}+1) \Big\},  \label{eq:H6}
\end{equation}
 where ${\hat n_v}$ stands for the number of valence nucleons, and
 ${\hat T}$ stands for the total isospin.
 It is well known that realistic effective interactions have large and
 comparable values of the centroids $\overline{V}(ab:T=0)$.
 The expression (\ref{eq:H6}) with a large average value $k^0$
 (for instance, $k^0=1.44$ MeV for $^{56}$Ni) shows that the symmetry energy
 comes mainly from $H_{mp}^{T=0}$ with the $T(T+1)$ term \cite{Kaneko}.
 The monopole field in the form (\ref{eq:H6}) can be regarded
 as an additional term to the Hartree-Fock mean field, in a sense.
  However, the residual interaction $H_{res}$, which determines
 the microscopic structure, contributes significantly
 to the symmetry energy \cite{Kaneko}.
  The symmetry energy cannot be reduced to a simple mean field
 but is affected by dynamical interactions in the shell model.
 
\begin{figure}[b]
\begin{center}
\includegraphics[width=6.8cm,height=7.5cm]{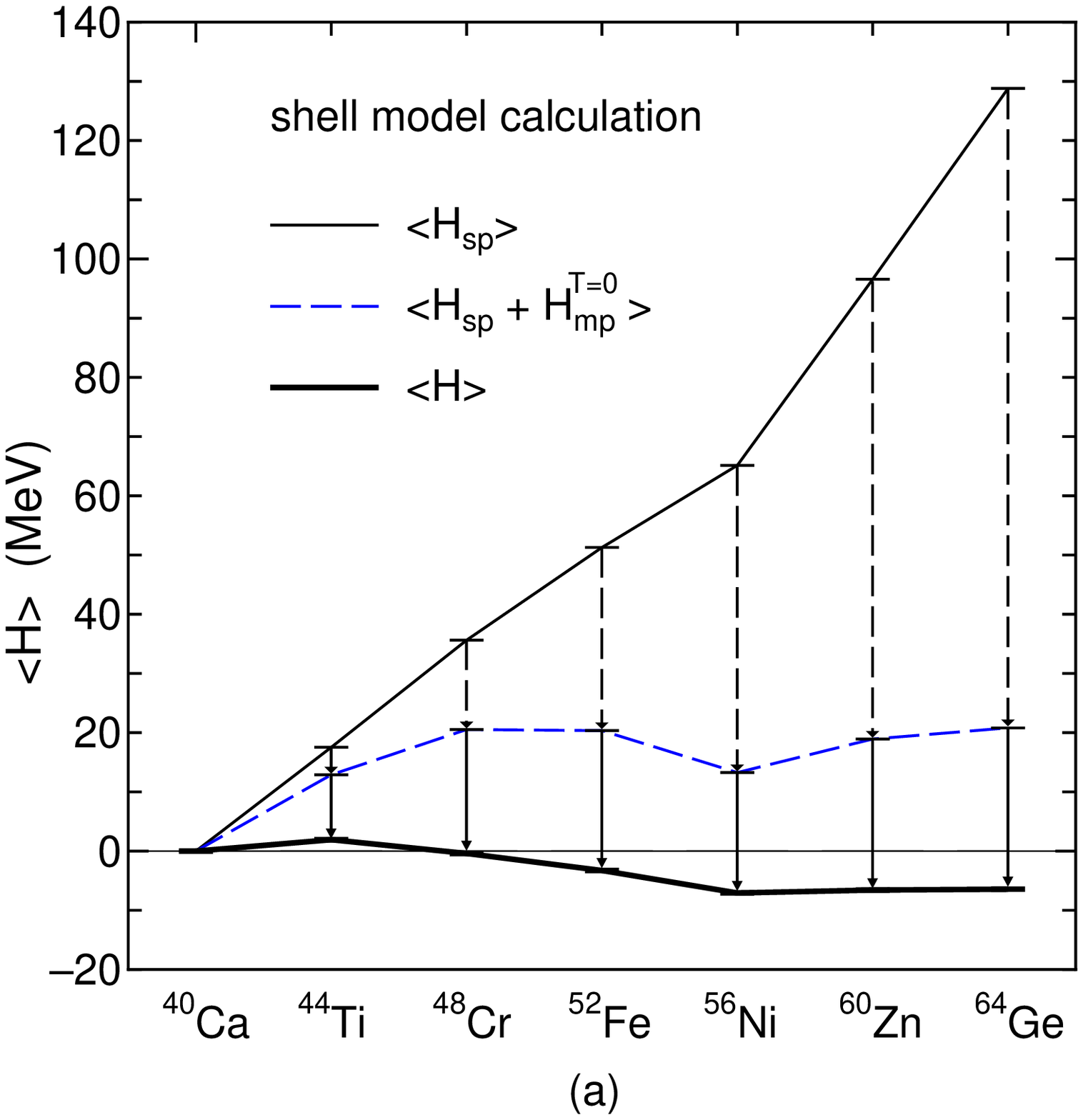}
\includegraphics[width=6.8cm,height=7.5cm]{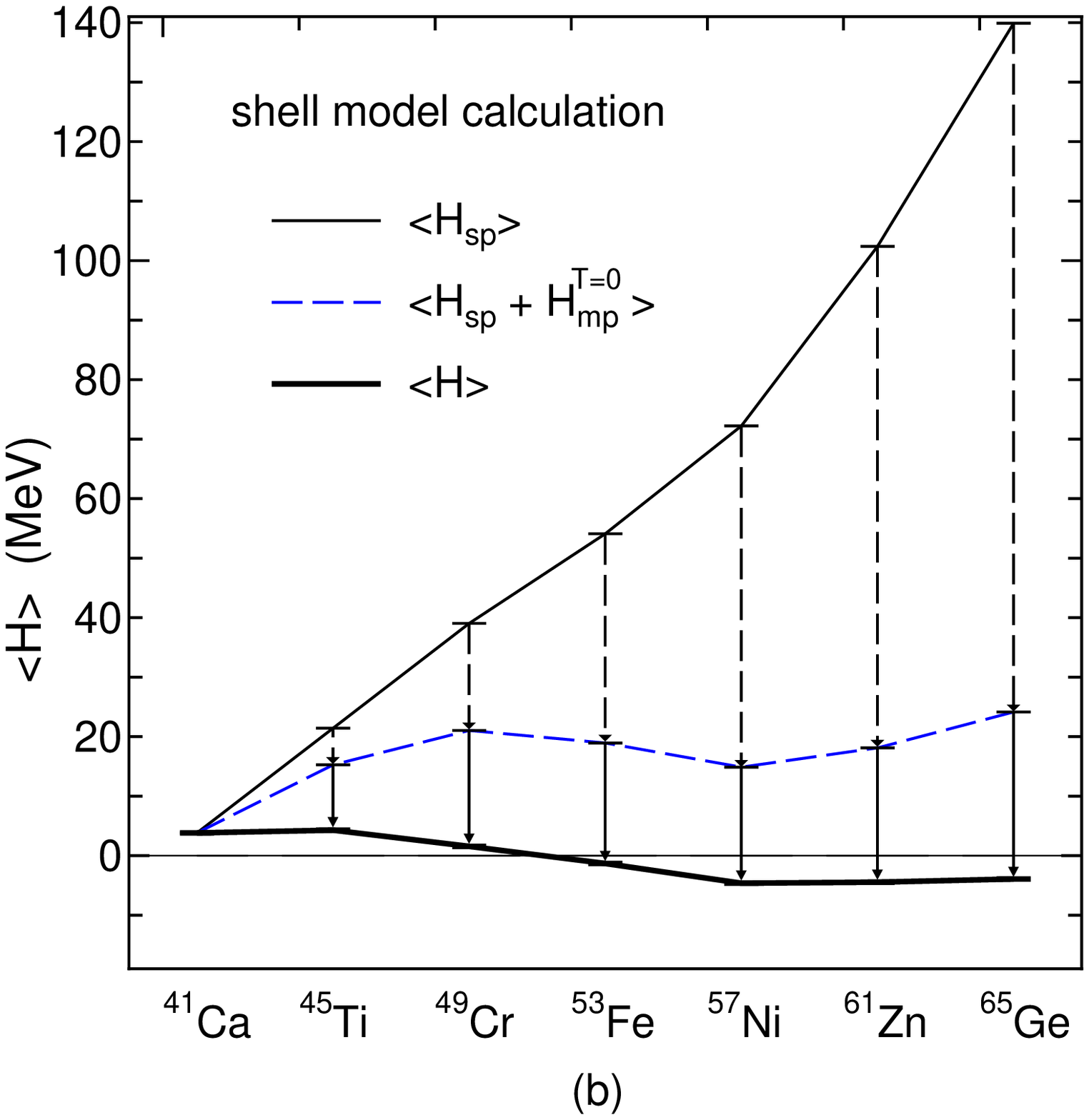}
  \caption{Expectation values of $H_{sp}$, $H_{sp}+H_{mp}^{T=0}$ and $H$
           for (a) even-even $N=Z$ nuclei with $A=40+m\alpha$ and (b)
           odd-$A$ nuclei with $A=40+m\alpha+1n$.
           The residual interaction energy $\langle H_{res}\rangle$
           is denoted by the solid-line arrows, and the monopole contribution
           $\langle H_{mp}^{T=0} \rangle$ is denoted by the dashed-line arrows.}
  \label{fig3}
\end{center}
\end{figure}

  We carried out numerical calculations using Mizusaki's code,
 \cite{Mizusaki,Mizusaki2} which makes large-scale shell model calculations
 possible by means of extrapolation.
 The calculated results for the $A=40+m\alpha$ nuclei from $^{44}$Ti to
 $^{64}$Ge are illustrated in Fig. \ref{fig3}(a),
 where $\langle H_{sp} \rangle$, $\langle H_{sp}+H_{mp}^{T=0} \rangle$
 and $\langle H \rangle$ denote their expectation values 
 for the ground state.  The behavior of the ground-state energy
 $\langle H \rangle$ corresponds well with that of the energy $E(A_0+m\alpha)$
 seen in Figs.~\ref{fig1} and \ref{fig2}.  Figure \ref{fig3}(a)
 shows that the energy $E(A_0+m\alpha)$ represents the ground-state energies
 of the even-even $N=Z$ nuclei and, moreover,
 that $E(A)$ hides significant correlations
 in the background. This figure supports the schematic explanation
 for the large energy gains of the $A_0+m\alpha$ nuclei in Fig. \ref{fig2}.
  Even if we regard the monopole field as a part of the mean field,
 the residual interaction energy $\langle H_{res} \rangle$ is still large
 in Fig. \ref{fig3}(a).
 The residual interaction energy $\langle H_{res} \rangle$ is essential
 for bringing the values of $E(A_0+m\alpha)$ ($ \approx \langle H \rangle$)
 close to the zero line.
 In $^{56}$Ni, for instance, $\langle H_{res} \rangle$ is approximately 20 MeV,
 which overwhelms the single-particle energy gap between
 $f_{7/2}$ and $p_{3/2}$. The closed-shell configuration $(f_{7/2})^{16}$
 does not exceed 68\% in the wave function of the ground state, according to
 Ref. 30). The new Tamm-Dancoff solution for the $J=T=0$
 four-particle excitation mode indicates that the ground state of $^{56}$Ni
 cannot be described within a perturbation expansion starting with
 the closed-shell configuration  \cite{Hase3}.
 This situation is called ``$\alpha$-like superfluidity" in the next section.
 The $^{56}$Ni nucleus can be regarded as a correlated state of
 valence nucleons outside the $^{40}$Ca core.
 Figure \ref{fig3}(a) shows the upward turn of $\langle H \rangle$
 from $^{56}$Ni to $^{60}$Zn resulting from an energy loss of four additional
 nucleons occupying the upper orbits beyond the semi-magic number $Z=N=28$. 
 The variation of $E(A_0+m\alpha)$ is therefore related to a shell effect
 as well as correlations.
 However, it should be noted that the position of
 $\langle H(^{60}\mbox{Zn}) \rangle$ in Fig. \ref{fig3}(a) depends on 
 significant collapse of the $^{56}$Ni core.

    Figure \ref{fig3}(b) displays the shell model results for odd-mass nuclei
 with $A_0+m\alpha+1n$. This figure is very similar to Fig. \ref{fig3}(a).
 The ground-state energy $\langle H \rangle$ exhibits a dip at $^{57}$Ni
 in Fig. \ref{fig3}(b), like the dip at $^{56}$Ni in Fig. \ref{fig3}(a).
 In the shell model results for $^{57}$Ni, the occupation probabilities
 of the respective orbits indicate strong correlations of valence nucleons
 and collapse of the $^{56}$Ni core, which is contrary to a simple picture
 of the $^{56}$Ni core plus one neutron. The single-particle energy gap
 between $f_{7/2}$ and $p_{3/2}$ is negligible as compared
 with the interaction energy, though the energy loss of four additional
  nucleons occupying the upper orbits ($p_{3/2}$, $p_{1/2}$, $f_{5/2}$)
 causes an upward turn of $\langle H \rangle$ from $^{57}$Ni to $^{61}$Zn.
 The present shell model also explains the behavior of the experimental energy
 $E(A_0+m\alpha+1n)$ shown in Fig. \ref{fig4}(b).
 Thus, Figs. \ref{fig3}(a) and \ref{fig3}(b) support our picture, which regards
 nuclei around $^{56}$Ni as correlated states of valence nucleons
 outside the $^{40}$Ca core.
 It is interesting that a $A_0+m\alpha+1n$ nucleus appears to be composed of
 an $A_0+m\alpha$ system and a last neutron with an effective single-particle
 energy. We can assume roughly the same correlations forming a common
 structure in the two nuclei with $A=A_0+m\alpha$ and $A=A_0+m\alpha+1n$.
 
   The correlations buried in the energies $E(A)$ of the $A_0+m\alpha$
 nuclei have been considered little in the framework of the mass formula.
 The inconspicuous values of $E(A_0+m\alpha)$, however, hide the $T=0$
 $2n-2p$ correlations which are stronger than the $T=1$ pair correlations,
 in the background. The energy $E(A_0+m\alpha)$ should be considered
 explicitly in the mass formulas. 
   Figure \ref{fig2} suggests a description of the energies $E(A)$
 of $N \ne Z$ nuclei with the indicators $\delta M$ and $\Delta$
 on the $T=0$ line connecting the $A_0+m\alpha$ nuclei. 
 We now note the importance of the $T=0$ line in Fig. \ref{fig1}. 
 Ignoring the variation of the $T=0$ line affects the masses of
 all $N \approx Z$ nuclei. 
  The values of $E(A_0+m\alpha)$ are not zero, and the substantial deviations
  should not be ignored.

\section{Fundamental $T=0$ $2n-2p$ correlated structure}

\begin{figure}[t]
\begin{center}
\includegraphics[width=6.8cm,height=7.4cm]{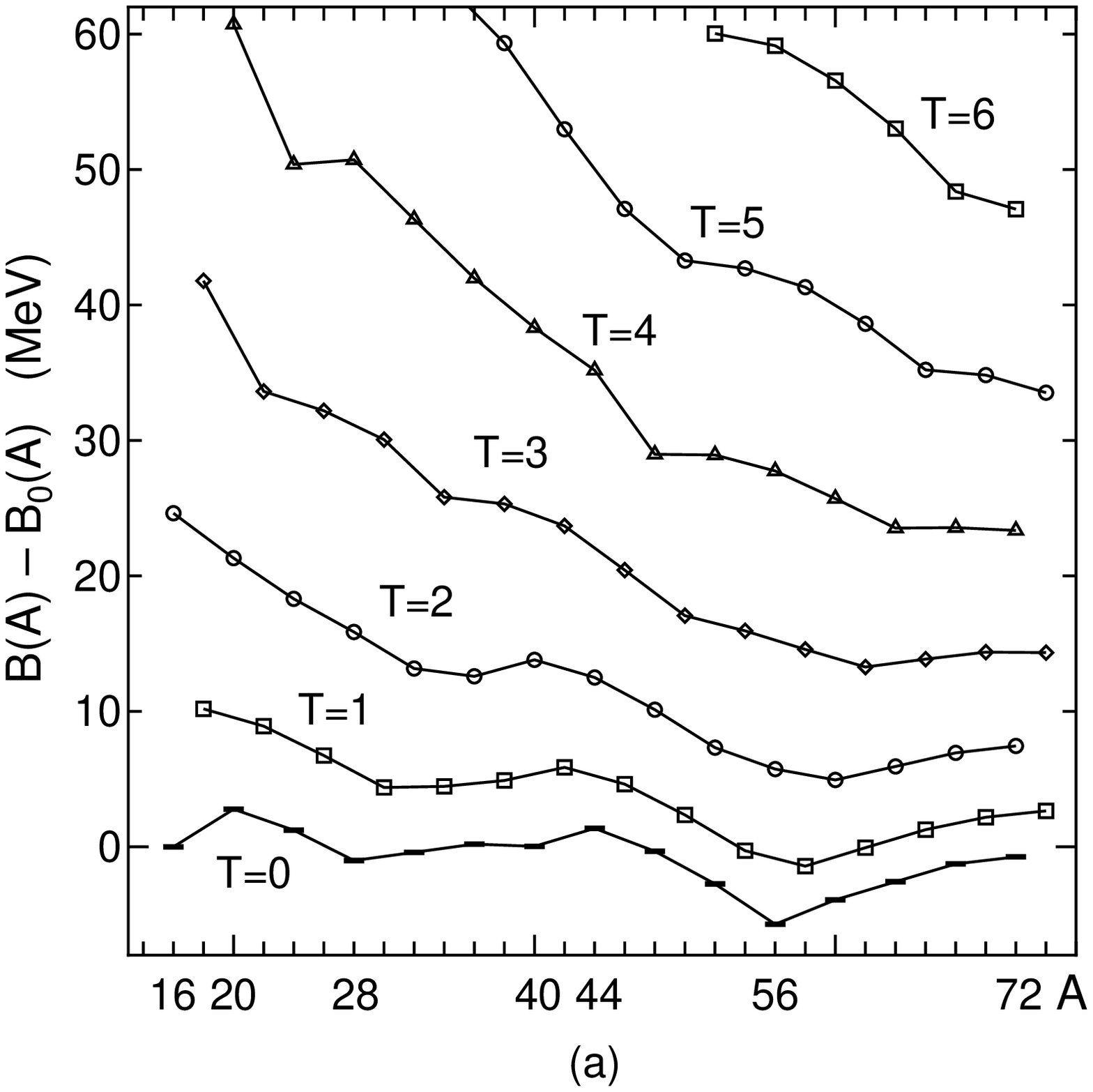}
\includegraphics[width=6.8cm,height=7.4cm]{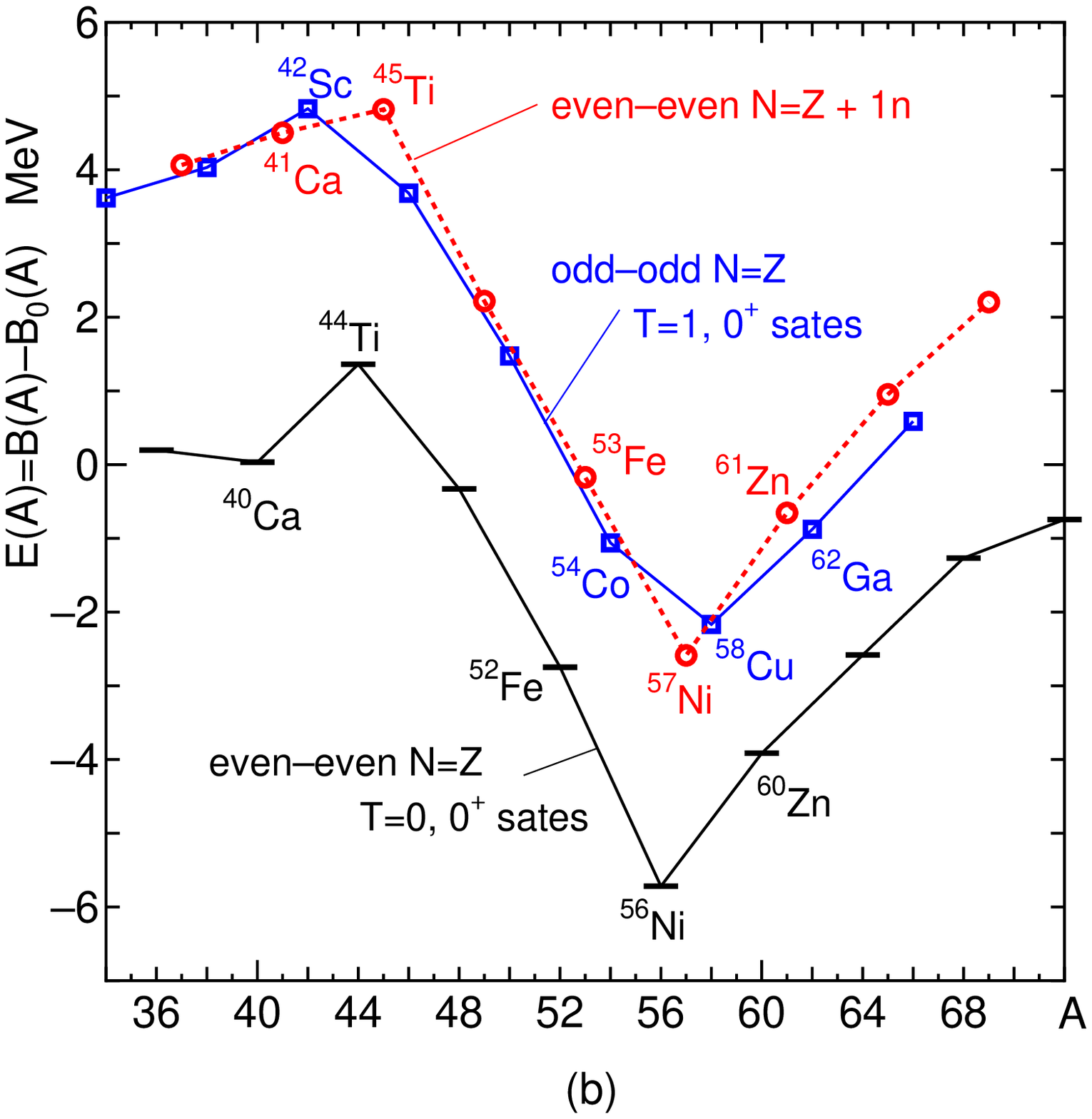}
  \caption{Experimental energies $E(A)$ defined in Eq. (\ref{eq:4})
           for (a) even-even nuclei with $N \ge Z$ and
           (b) nuclei with $A_0+m\alpha$, $A_0+m\alpha+1n$ and
           $A_0+m\alpha+1n1p$ around $^{56}$Ni.}
  \label{fig4}
\end{center}
\end{figure}

   We have seen significant correlations buried in $E(A)$ of $A_0+m\alpha$
 and $A_0+m\alpha+1n$ nuclei.
 What is the nature of the correlations in other $N \neq Z$ nuclei?
 To answer this, let us consider Fig.~\ref{fig4}(a),
  in which experimental values of $E(A)$ are plotted
 for even-even nuclei with $N \ge Z$. 
 Even-even nuclei with $N > Z$ can be classified according to the number of
 neutron pairs added to the $A_0+m\alpha$ systems,
 such as $A=A_0+m\alpha +2n$, $A=A_0+m\alpha +4n$, $\cdots$,
 and each series of them with increasing $m$ has the same $T$.
  Figure \ref{fig4}(a) indicates a parallelism
  with regard to $E(A)$ between even-even nuclei with $T>0$ and $T=0$.
   More precisely, every $T$ line connecting a series of nuclei
 is parallel to the $T=0$ line of the $A_0+m\alpha$ nuclei. 
 The same parallelism is seen in the experimental values of $E(A)$
 for even-even $N \le Z$ nuclei
 ($A=A_0+m\alpha$, $A=A_0+m\alpha +2p$, $A=A_0+m\alpha +4p$, $\cdots$),
 though they are omitted in Fig. \ref{fig4}(a) for simplicity.
 The experimental energy $E(A)$ for the $T=1$, $0^+$ states of odd-odd $N=Z$
 nuclei also varies in a manner parallel to $E(A_0+m\alpha)$,
  as shown in Fig. \ref{fig4}(b).
 Realistic shell model calculations faithfully reproduce the parallelism of
 the experimental $E(A)$ in Fig. \ref{fig4}.
 This parallelism is expressed in terms of the symmetry energy
 in ordinary mass formulas.
 
   It should, however, be noted that the characteristic behavior of
 $E(A_0+m\alpha)$, which hides important correlations and resulting structure,
 appears in the $T>0$ lines of other nuclei. 
  The parallel variations of $E(A)$ suggest the existence of a common structure
 formed in nuclei with $A=A_0+m\alpha$, $A=A_0+m\alpha +2n(2p)$,
 $A=A_0+m\alpha +4n(4p)$, etc., and also in $A_0+m\alpha +1n1p$ nuclei
 with $T=1$. 
 It is notable in Fig. \ref{fig4}(b) that parallel variation of $E(A)$
 appears also in the odd-$A$ nuclei with $A=A_0+m\alpha +1n$, which is
 seen in the shell model results of Fig. \ref{fig3}. The energy difference
 $E(A_0+m\alpha+1n)-E(A_0+m\alpha)$ is related to the pairing energy
 in the ordinary mass formulas.
 This parallelism again suggests a common structure in the two nuclei
 with $A=A_0+m\alpha$ and $A=A_0+m\alpha +1n$.
 This common structure is probably the $T=0$ $2n-2p$ correlated
 structure of the $A_0+m\alpha$ nuclei.
 
\subsection{Multi-quartet structure of even-even $N=Z$ nuclei}

  The study of nuclear structure has clarified the importance of
 the $T=0$ $2n2p$ correlations in $N \approx Z$ nuclei.
  Recall that the core plus $\alpha$ cluster (two $\alpha$ cluster)
 model accurately describes $^{20}$Ne, $^{24}$Mg, $^{44}$Ti and even $^{48}$Cr.
  In a simplified picture ignoring the spatial correlations of $\alpha$,
  the ground states of
  $A_0+m\alpha$ nuclei can be approximated in the following way
  \cite{Hase1}:
\begin{equation}
  |\Phi_0(A_0+m\alpha) \rangle \approx \frac{1}{\sqrt{N(A_0+m\alpha)}}
            (\alpha_{J=T=0}^\dagger)^m |A_0 \rangle,         \label{eq:6}
\end{equation}
 where $\alpha_{J=T=0}^\dagger$ consists of a linear combination 
 of four valence nucleons
 $(c^\dagger)^4_{J=T=0}$ determined by the Tamm-Dancoff equation
 $[H, \alpha_{J=T=0}^\dagger ] \approx e_\alpha \alpha_{J=T=0}^\dagger$,
  $N(A_0+m\alpha)$ is a normalization constant,
 and $|A_0 \rangle$ denotes the doubly-closed-shell core.
 When the Hamiltonian has only two-body interactions,
 the energy $E(A_0+m\alpha)$ can be calculated as
\begin{eqnarray}
 & {} &  E(A_0+m\alpha) = m \frac{\langle A_0| (\alpha_0)^m (\alpha_{J=T=0}^\dagger)^{m-1}
  [H, \alpha_{J=T=0}^\dagger ]|A_0 \rangle}{N(A_0+m\alpha)} \nonumber \\ 
 & {} &    + \frac{1}{2} m(m-1)
   \frac{\langle A_0| (\alpha_0)^m (\alpha_{J=T=0}^\dagger)^{m-2}
   [[H, \alpha_{J=T=0}^\dagger ], \alpha_{J=T=0}^\dagger ]|A_0 \rangle}{N(A_0+m\alpha)}. 
       \label{eq:7}
\end{eqnarray}
 The microscopic calculations of Eq. (\ref{eq:7}) in Ref. 25)
 approximately reproduce the experimental energies $E_{shl}(A_0+m\alpha)$
 [and hence $E(A_0+m\alpha)$] for $^{44}$Ti, $^{48}$Cr, $^{52}$Fe and $^{56}$Ni.
 Therefore, the $A_0+m\alpha$ nuclei have a multi-quartet structure
 approximated by Eq. (\ref{eq:6}), at least up to $^{56}$Ni.
 The characteristic behavior of $E(A_0+m\alpha)$ in Fig. \ref{fig1}
 indicates the leading role of the multi-quartet structure.
 The shell model results in $\S$2.3 allow us to imagine the multi-quartet
 structure for the $A_0+m\alpha$ nuclei beyond $^{56}$Ni.

   In order to get a simple formula, let us transfer the fermion equation
 (\ref{eq:7}) into an interacting $\alpha$-boson model,
\begin{equation}
  E_{IBM}(A_0+m\alpha) = m e_\alpha - \frac{1}{2} m(m-1) G_{\alpha \alpha},
           \label{eq:8}
\end{equation}
 where $e_\alpha$ is the energy of the $\alpha$-boson and $G_{\alpha \alpha}$
 denotes the interaction between the $\alpha$-bosons.  Because the $\alpha$-boson
 is regarded as being mapped from the four correlated fermions, the interaction
 strength $G_{\alpha \alpha}$ should reflect the Pauli principle.
 The stable doubly-closed-shell nuclei $^{16}$O and $^{40}$Ca suggest
 that the four correlated nucleons are mainly in one major shell \cite{Tomo}.
 As the number of $\alpha_{J=T=0}^\dagger$ increases in the major shell,
 the Pauli principle applied to $\alpha_{J=T=0}^\dagger$ must restrict
 the degrees of freedom for $\alpha_{J=T=0}^\dagger$.
 Let us take the Pauli principle effect into account by expressing
 the interaction strength $G_{\alpha \alpha}$ in the form of a decreasing
 function of $m$ (the number of $\alpha_{J=T=0}^\dagger$).  The simplest way
 to do this is
 to approximate the decline function by a linear function of $m$, such as
\begin{equation}
  G_{\alpha \alpha} = g_{\alpha \alpha} \{ 1 - C_\alpha (m-1) \}.   \label{eq:9}
\end{equation}
 The factor $C_\alpha$ represents something like the scale of the subspace
 $\{(\alpha_{J=T=0}^\dagger)^m|A_0 \rangle \}$, depending on the shell structure.
 This interacting $\alpha$-boson model can reproduce the experimental values
 of $E(A_0+m\alpha)$ up to $^{56}$Ni, as shown in Fig. \ref{fig1},
 where the values denoted by the open squares are obtained with the parameter
 values
 $e_\alpha=2.75$, $g_{\alpha \alpha}=5.3$, $C_\alpha=0.21$ in MeV
 for $^{20}$Ne to $^{28}$Si and $e_\alpha=1.35$, $g_{\alpha \alpha}=3.75$,
 $C_\alpha=0.18$ in MeV for $^{44}$Ti to $^{56}$Ni.
  The interacting $\alpha$-boson model describes the peaks
 at $A=A_0+\alpha$ ($^{20}$Ne and $^{44}$Ti) and the decline toward $^{28}$Si
 and $^{56}$Ni. 

   The most important point here is that the interaction between the composite
 quartets $\alpha_{J=T=0}^\dagger$ is attractive and quite strong.
 Other composite fermion units, like Cooper pairs and vibrational phonons
 with $J=2$ and $J=3$ in nuclear physics, have repulsive interactions
 between them and do not actually have a boson-like property, because of
 the Pauli principle.  Only the $\alpha$-like quartet with $J=T=0$ has
 the possibility to resemble a boson, because of a special mechanism
 in couplings of spin and isospin.
  The large energy gain due to the strong $T=0$ $2n-2p$ correlations
 and the attractive interaction between the $\alpha$-like quartets cause
 the effect of the (collective) $T=0$ $2n-2p$ correlations
 on the nuclear mass to be rather inconspicuous.
 
    The interacting $\alpha$-boson model (\ref{eq:8}) with (\ref{eq:9}) 
 roughly reproduces the energies $E(^{32}$S$)$ and $E(^{60}$Zn$)$ given in Fig.
 \ref{fig1}.  The shell model calculation in $\S$2.3, however, shows
 that the upward turn to $^{60}$Zn in the graph of $E(A_0+m\alpha)$ is
 due to a shell effect.  The formula (\ref{eq:8}) cannot be applied
 to the regions $32<A<40$ and $60<A<80$, because the expression (\ref{eq:9})
 is not valid there. A hole picture is probably suitable for these latter halves
 of the $sd$ and $pf$ shells. Then we obtain the same type of states as in Eq.
 (\ref{eq:6}) by replacing $\alpha_{J=T=0}^\dagger$ with a linear combination
 of four holes $(c_h^\dagger)^4_{J=T=0}$. The corresponding boson picture,
 the interacting $\alpha$-hole boson model, may give a formula similar to
 (\ref{eq:8}).
 It is, however, difficult to obtain a simple formula that reproduces 
 the variation of $E(A_0+m\alpha)$ including a shell effect in the entire region
 of $N=Z$ nuclei.  We therefore abandon this problem and instead adopt
 the experimental values given in Fig. \ref{fig1} for the energy
 $E(A_0+m\alpha)$ in this paper.

  The approximation (\ref{eq:6}) is very simplified in comparison with
 the realistic shell model.  Adding other collective modes
 of the $T=0$ $2n-2p$ correlations with $J>0$ is necessary to better reproduce
 the variation of $E(A_0+m\alpha)$ for the $f_{7/2}$ shell nuclei \cite{Hase1B}.
 We should consider the $T=0$ $2n-2p$ correlations as correlations of collective
 $T=0$ $2n-2p$ quartets with various $J$ in an improved approximation.
 In fact, although we usually imagine the multi $J$$=$$0$ pair structure
 for the $T=1$ pair correlated state, a realistic shell model wave function
 includes components of various $J>0$ pairs in nuclear physics.
 We use the term ``$T=0$ $2n-2p$ correlations" in such a broader sense.
 In the following sections, we express the $T=0$ $2n-2p$ correlated states as
 \begin{equation}
  |\Phi_0(A_0+m\alpha) \rangle \propto (\alpha_{T=0}^\dagger)^m |A_0 \rangle,                 \label{eq:10}
\end{equation}
 where $\alpha_{T=0}^\dagger$ represents a quartet of $T=0$ $2n-2p$ correlated
 nucleons or holes.
 We formally use the expression (\ref{eq:10}) also in the hole regions,
 $32<A<40$ and $60<A<80$.

\subsection{Superfluid state induced by the $T=0$ $2n-2p$ correlations}
 
   In Fig. \ref{fig1}, the line connecting the energies $E(A_0+m\alpha)$
 of the $T=0$ nuclei plays an important role in the mass formula,
 because the energies of the other nuclei with $T>0$ are measured from this line.
 For $E(A_0+m\alpha)$, we provisionally use the experimental values
 evaluated from $B(A_0+m\alpha)-B_0(A_0+m\alpha)$, as mentioned above.
   Let us extrapolate the line for $E(A_0+m\alpha)$
 to nuclei with $A \neq A_0+m\alpha$ as follows:
\begin{eqnarray}
  E_{T=0}(A_0+m\alpha+2) & = & (E(A_0+m\alpha) + E(A_0+m\alpha+\alpha)) /2, \nonumber \\
  E_{T=0}(A_0+m\alpha+1) & = & (E(A_0+m\alpha) + E_{T=0}(A_0+m\alpha+2)) /2, \nonumber \\
  E_{T=0}(A_0+m\alpha+3) & = & (E_{T=0}(A_0+m\alpha+2) + E(A_0+m\alpha+\alpha)) /2.
    \label{eq:11}
\end{eqnarray}
 These equations define the $T=0$ plane as the base level of energy
 in the mass table.  At this stage, the binding energy is written
\begin{equation}
  B(A) = B_0(A) + E_{T=0}(A) + W(A).        \label{eq:12}
\end{equation}
 The pairing energy, symmetry energy, Wigner energy and a correction
 for odd-odd nuclei are included in the residual energy $W(A)$
 in Eq. (\ref{eq:12}).
 
   The $T=0$ $2n-2p$ correlations are related to
 the ``$\alpha$-like superfluidity" proposed in Ref. 7),
 where, by analogy to pairing superfluidity, $\alpha$-like superfluidity
 is indicated by the following mass difference corresponding to the odd-even
 mass difference $\Delta$:
\begin{eqnarray}
  \delta M(A_0+m\alpha+2) & = & (B(A_0+m\alpha+2n)+B(A_0+m\alpha+2p))/2   \nonumber \\
               & - & (B(A_0+m\alpha)+B(A_0+m\alpha+\alpha))/2.    \label{eq:13}
\end{eqnarray}
 This quantity is called the ODD-EVEN mass difference in $\S$2.2.
 The average energy of the $A_0+m\alpha+2n$ and $A_0+m\alpha+2p$ nuclei
 is used so as to remove the Coulomb energy effect.  Although the
 $\alpha$-like superfluidity in heavy nuclei is discussed in Ref. 7),
 we are concerned with $N \approx Z$ nuclei, for which the isospin is
 a good quantum number.  In place of Eq. (\ref{eq:13}), we define the
 following quantity as an indicator of the $\alpha$-like superfluidity:
\begin{eqnarray}
  \delta W(A_0+m\alpha+2:T=1,K) & = & W(A_0+m\alpha+2:T=1,K)   \nonumber \\
    & - & (W(A_0+m\alpha)+W(A_0+m\alpha+\alpha))/2,    \label{eq:14}
\end{eqnarray}
 where $K=1$, 0 and $-1$ correspond to the $T=1$ states of the $A_0+m\alpha+2n$,
 $A_0+m\alpha+1n1p$ and $A_0+m\alpha+2p$ nuclei, respectively.  The residual
 energy $W(A)$ is measured from the $T=0$ plane [$E_{T=0}(A)$],
 and hence we have $W(A_0+m\alpha)=W(A_0+m\alpha+\alpha)=0$.
   It is thus seen that $W(A_0+m\alpha+2:T=1,K)$
 is identically the ODD-EVEN mass difference for $\alpha$-like superfluidity,
\begin{equation}
  \delta W(A_0+m\alpha+2:T=1,K) = W(A_0+m\alpha+2:T=1,K).    \label{eq:15}
\end{equation}
 The average of $W(A_0+m\alpha+2:T=1,K=1)$ and $W(A_0+m\alpha+2:T=1,K=-1)$ 
 corresponds with $\delta M(A_0+m\alpha+2)$ in Eq. (\ref{eq:13}),
 which represents the $n-p$ (mainly $T=0$) interaction energy between $2n$
 and $2p$ in an $A_0+m\alpha+\alpha$ nucleus \cite{Kaneko}.

\begin{figure}[t]
\begin{center}
\includegraphics[width=7cm,height=7cm]{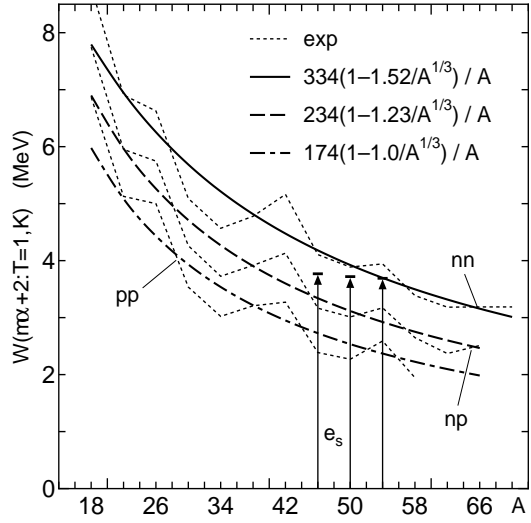}
  \caption{The energy $W(A_0+m\alpha+2:T=1,K)$
           for $A_0+m\alpha+2$ nuclei with $T=1$. This quantity represents
           the ODD-EVEN mass difference for $\alpha$-like superfluidity.}
  \label{fig5}
\end{center}
\end{figure}

   The values of $W(A_0+m\alpha+2:T=1,K)$ for the $A_0+m\alpha+2n$,
 $A_0+m\alpha+1n1p$ and $A_0+m\alpha+2p$ nuclei are plotted
 by the dotted curves in Fig. \ref{fig5}.
 Figure \ref{fig5} shows that the ODD-EVEN mass difference for the $T=0$ $2n-2p$
 correlations is larger than the odd-even mass difference $\Delta$
 for the pairing superfluidity. 
 The shell model calculation in $\S$2.3 shows
  that there is a significant contribution of
 the monopole field $H_{mp}^{T=0}$ to the ODD-EVEN mass difference
 [in Eq. (\ref{eq:14})]. The $H_{mp}^{T=0}$ contribution is $3k^0/2$.
 For $A\approx 58$, for instance, the value is about 2.14 MeV,
 while $W(A=58:T=1,K=1) \approx 3.4$ MeV.
 The $H_{res}$ contribution to the ODD-EVEN mass difference is about 1.25 MeV,
 which is comparable to the odd-even mass difference
 $\Delta \approx 1.34$ MeV near $A=58$.
 It should be noted here that the $T=1$ pair correlations of neutron and proton
 pairs joining in the formation of the $T=0$ $2n-2p$ quartet are not included
 in the ODD-EVEN mass difference. We can say that the strong $T=0$ $2n-2p$
 correlations cause a superfluid state, like the pairing superfluid state,
 as claimed by Gambhir {\it et al.}\cite{Gamb}
  It is notable that the ODD-EVEN mass difference for $N \approx Z$ nuclei
 is larger than that for $N>Z$ nuclei, with regard to which the term
 ``$\alpha$-superfluidity" was first used \cite{Gamb}.
 We call the strongly correlated state an ``$\alpha$-like superfluid state".
 As discussed in the previous subsection, the $\alpha$-like superfluid state
 has the multi-quartet structure (\ref{eq:6}) in the $A_0+m\alpha$ nuclei,
 at least up to $^{56}$Ni.

 Figure \ref{fig5} displays the systematic differences among the $A_0+m\alpha+2n$,
 $A_0+m\alpha+1n1p$ and $A_0+m\alpha+2p$ nuclei. This indicates that the effect
 of the Coulomb interaction remains after subtracting the Coulomb energy
 term $a_C Z^2/A^{1/3}$.  It may be necessary for a practical mass formula
 to add some correction terms in order to remove the differences
 between the states with different $K$.  In fact, modern mass formulas do have
 such correction terms.  However, we leave this problem and
 employ different parameters for neutrons and protons in this paper,
 where we aim to explain our basic idea.
 
\subsection{Bogoliubov transformation for the $\alpha$-like superfluid state}
 
   We consider $A_0+m\alpha+2$ nuclei with $T=1$,
where structure is roughly expressed as 
\begin{eqnarray}
 & {} & |A=A_0+m\alpha+2:T=1,K \rangle \propto S_K^\dagger
        (\alpha_{T=0}^\dagger)^m |A_0 \rangle ,  \label{eq:16} \\
 & {} & S_K^\dagger \propto (c^\dagger c^\dagger)_{J=0,T=1,K} . \nonumber
\end{eqnarray}
 The large values of $W(A_0+m\alpha+2:1K)$ allow us to regard
 the state
  $|\Phi_0(A_0+m\alpha) \rangle $ as an $\alpha$-like superfluid state
 $(\alpha_{T=0}^\dagger)^m |A_0 \rangle$.  After a kind of the Bogoliubov
 transformation, the $\alpha$-like superfluid state is the vacuum state
 $|0(\alpha) \rangle$ for a quasi-pair $\mbox{\boldmath $S$}_K^\dagger$,
 which is transformed from $S_K^\dagger$. In this picture, the state
 (\ref{eq:16}) is regarded as a quasi-pair state, like the quasi-particle
 state in the BCS theory,
\begin{equation}
 |A=A_0+m\alpha+2:1,K \rangle = \mbox{\boldmath $S$}_K^\dagger
              |0(\alpha) \rangle . \label{eq:17}
\end{equation}
 Measuring the energy $W(A)$ from the $T=0$ plane [$E_{T=0}(A)$]
 defined in Eq. (\ref{eq:11}) corresponds to the above transformation
 for the wave functions.  The energy $W(A_0+m\alpha+2:1K)$ is the energy
 of the quasi-pair $\mbox{\boldmath $S$}_K^\dagger$. 
 This discussion is parallel to the quasi-particle picture
 concerning the pairing energy, as seen below.

   The above transformation is, in fact, difficult to carry out
  for the four composite fermions $\alpha_{T=0}^\dagger$.
    Instead, we illustrate our plan using the interacting boson model (IBM),
     as done by Gambhir {\it et al.} \cite{Gamb}.
 The IBM for $N \approx Z$ nuclei is called the IBM3.  The IBM3 Hamiltonian
 is expressed in terms of the $s$ boson ($J=0$) and $d$ boson ($J=2$)
 with $T=1$ \cite{Thompson,Hase4},
\begin{equation}
  s_K^\dagger = s_{J=0,T=1,K}^\dagger , \quad
  d_{MK}^\dagger = d_{2M1K}^\dagger.   \label{eq:18}
\end{equation}
 The $sd$ boson image of $\alpha_{J=T=0}^\dagger$ is given by
\begin{equation}
  \alpha_{J=T=0}^\dagger \Rightarrow x (s^\dagger s^\dagger)_{J=T=0}
   + \sqrt{1-x^2} (d^\dagger d^\dagger)_{J=T=0}.   \label{eq:19}
\end{equation}
 For the $\alpha$-like superfluid state, the quasi $s$ and $d$ bosons
 ($\mbox{\boldmath $s$}_K$, $\mbox{\boldmath $d$}_{MK}$) are introduced
 through the Bogoliubov transformation
\begin{eqnarray}
     s_K^\dagger &=& U_s \mbox{\boldmath $s$}_K^\dagger
                 + V_s (-)^{1-K} \mbox{\boldmath $s$}_{-K} ,   \nonumber \\
  d_{MK}^\dagger &=& U_d \mbox{\boldmath $d$}_{MK}^\dagger
                 + V_d (-)^{2-M}(-)^{1-K}\mbox{\boldmath $d$}_{-M-K}. \label{eq:20}
\end{eqnarray}
 Here, we have $U_i^2 - V_i^2 =1$ ($i=s$, $d$).
 We have a boson-type gap equation and can calculate the quasi-boson
 energies $e_s$ and $e_d$ using an appropriate IBM3 Hamiltonian.
 
 In this quasi-boson picture, the quasi-pair state (\ref{eq:17})
 of the $A_0+m\alpha+2$ nuclei is written
\begin{equation}
 |A=A_0+m\alpha+2:1,K \rangle \Rightarrow \mbox{\boldmath $s$}_K^\dagger |0(\alpha)),
     \label{eq:21}
\end{equation}
 where the $\alpha$-like superfluid vacuum state is replaced by that
 for the quasi-bosons ($\mbox{\boldmath $s$}_K$, $\mbox{\boldmath $d$}_{MK}$). 
 There is a well-defined IBM3 Hamiltonian for the $f_{7/2}$ shell nuclei
 \cite{Thompson}.
 Using the IBM3 Hamiltonian, we evaluated the quasi-boson energy $e_s$,
 which should be equal to $W(A_0+m\alpha+2:T=1)$ given in Eq. (\ref{eq:15}).
 The calculated values of $e_s$ for $^{46}$Ti, $^{50}$Cr and $^{54}$Fe
 are plotted in Fig. \ref{fig5}.  
 It is seen that the quasi-boson energies $e_s$ accurately reproduce
 the experimental values of $W(A_0+m\alpha+2n)$ which is least sensitive
 to the Coulomb interaction effect. 
 This success supports our interpretation of the $\alpha$-like superfluidity
 of $A_0+m\alpha+2$ nuclei.

\section{Multi-pair structure on the base of $\alpha$-like superfluidity}

   Because the picture of $\alpha$-like superfluidity is good, the $J=0$
ground states of even-even nuclei can be approximated by
\begin{eqnarray}
 |A=A_0+m\alpha+2l:T=l \rangle & \propto & (S^\dagger)^l
        |\Phi_0(A_0+m\alpha) \rangle ,  \nonumber \\
     & \Rightarrow & (\mbox{\boldmath $s$}^\dagger)^l |0(\alpha)). \label{eq:22}
\end{eqnarray}
 Similar wave functions are considered in the microscopic derivation
 of a mass formula \cite{Zuker}.  Now we have reached the second stage,
 which can be compared with the first stage considering the
 multi-quartet state $(\alpha_{T=0}^\dagger)^m|A_0 \rangle$.
 We have another interacting boson picture for the Cooper pair,
\begin{equation}
 W(A_0+m\alpha+2l:T=l) = l e_s + \frac{1}{2} l(l-1) g_{ss}. \label{eq:23}
\end{equation}
 The interaction between the Cooper pairs (like-nucleon pairs)
 is repulsive because of the Pauli principle.
 The repulsive interaction between the quasi-$s$-bosons gives a quadratic
 increase of the mass, depending on the boson number $l$.
  The quasi-$s$-boson $\mbox{\boldmath $s$}_K^\dagger$ increases the isospin
 of the state by 1, and the number of $\mbox{\boldmath $s$}_K^\dagger$
 can be replaced with the isospin $T$ in Eq. (\ref{eq:23}).
 Let us write Eq. (\ref{eq:23}) in the ordinary form
\begin{eqnarray}
  W(A_0+m\alpha+2l:T=l) & = & a_{sym} T^2 + b_{Wig} T ,  \label{eq:24} \\
  a_{sym} & = & \frac{1}{2}g_{ss}, \quad b_{Wig}=e_s-\frac{1}{2}g_{ss}.
                      \label{eq:25}
\end{eqnarray}
 The first term here is called the symmetry energy, and the second term is called
 the Wigner energy in the mass formulas.  Our interacting boson picture
 for the multi-pair states explains the structural origins
 of the symmetry energy and Wigner energy.
 
\begin{figure}[b]
\begin{center}
\includegraphics[width=8cm,height=8cm]{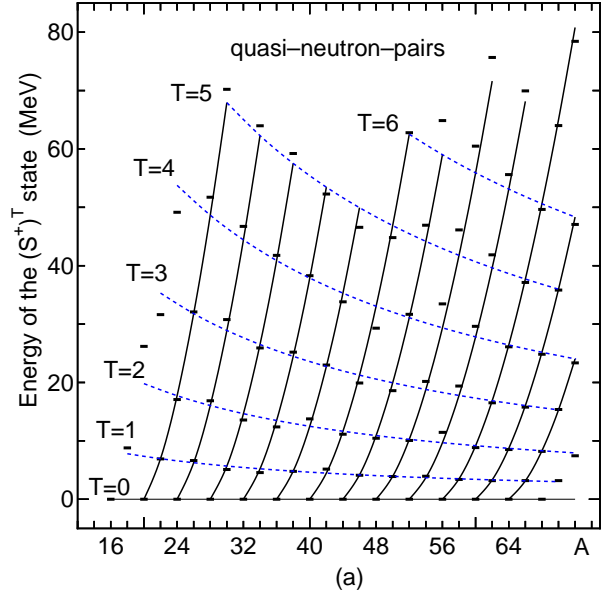}
  \caption{Energies $W(m\alpha + 2ln:T=l)$ of multi-quasi-neutron-pair states
           $(S_n^\dagger)^{l=T}|\Phi_0(m\alpha) \rangle$.
           Experimental values (flat dots) are compared with the theoretical
           values obtained with the parameters (\ref{eq:26})
           (which are at the intersections of the solid and dotted curves).}
  \label{fig6}
\end{center}
\end{figure}
 
\begin{figure}
\begin{center}
\includegraphics[width=8cm,height=5.71cm]{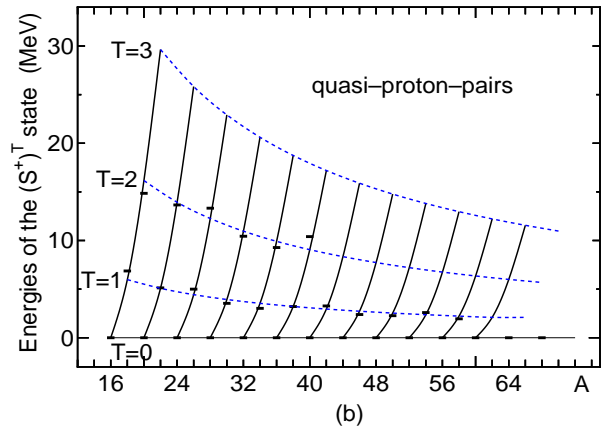}
  \caption{Energies $W(m\alpha + 2lp:T=l)$ of multi-quasi-proton-pair states
           $(S_p^\dagger)^{l=T}|\Phi_0(m\alpha) \rangle$, 
           shown in the same manner as Fig. \ref{fig5}.
           The theoretical values (at the intersections of the two curves)
           are obtained with the parameters (\ref{eq:27}).}
  \label{fig7}
\end{center}
\end{figure}

   Figures \ref{fig6} and \ref{fig7} display the experimental energies of
 the multi-quasi-pair states (\ref{eq:22}), $W(A_0+m\alpha+2ln)$ and
 $W(A_0+m\alpha+2lp)$. We can fix the parameter values $e_s$ and $g_{ss}$
 in the approximation (\ref{eq:23}) [$a_{sym}$ and $b_{Wig}$ in Eq.
 (\ref{eq:24})] from the experimental values of $W(A_0+m\alpha+2l:T=l)$.
  They can be expressed in the same form as that determined 
 in the microscopic mass formula \cite{Duflo}.
 The parameters, which are fixed separately for neutrons and protons, are
\begin{eqnarray}
  a_{sym}^{(n)} & = & 116 (1-1.52/A^{1/3})/A,  \nonumber \\
  b_{Wig}^{(n)} & = & 218 (1-1.52/A^{1/3})/A,  \label{eq:26} \\
  a_{sym}^{(p)} & = & 82 (1-1.0/A^{1/3})/A,  \nonumber \\
  b_{Wig}^{(p)} & = & 92 (1-1.0/A^{1/3})/A.  \label{eq:27}
\end{eqnarray}
 The quasi-$s$-boson energies $e_s^{(n)}$ and $e_s^{(p)}$ are plotted
 by the solid and dash-dot curves in Fig. \ref{fig5} [where a curve fitted
 to $W(A_0+m\alpha+1n1p)$ is also shown].  We see in Figs. \ref{fig6} and
 \ref{fig7} that the approximation (\ref{eq:24}) with the parameters
 (\ref{eq:26}) and (\ref{eq:27}) is very good, and hence our interacting
 boson picture for Cooper pairs is also good.  From the parameters
 (\ref{eq:26}) and (\ref{eq:27}), the coefficient of the Wigner energy
 is larger than that of the symmetry energy
 ($b_{Wig}^{(n)} \approx 1.88 a_{sym}^{(n)})$ for neutrons and
 $b_{Wig}^{(p)} \approx a_{sym}^{(p)}$ for protons. 
 The symmetry energy coefficient $a_{sym}^{(n)}$ is nearly equal to
 that determined in Ref. 4).
 The values of these parameters depend on the manner of evaluating
 the Coulomb energy.

  We point out that the quasi-$s$-boson energy is, for instance,
 $e_s^{(n)} \approx 6.4$ MeV for $A=20$ and $e_s^{(n)} \approx 4.3$ MeV
 for $A=44$, as obtained from Fig.~\ref{fig5}. [The interaction energy between
 the quasi-$s$-bosons is $g_{ss}^{(n)} \approx 5.1$ MeV for $A=20$
 and $g_{ss}^{(n)} \approx 3.0$ MeV for $A=44$, from Eq. (\ref{eq:26}).]
 The quasi-$s$-boson energy $e_s$ is larger than the $\alpha$-boson
 energy $e_\alpha$ ($e_\alpha =1.35$ MeV for the $sd$ shell nuclei and
 $e_\alpha =2.75$ MeV for the $pf$ shell nuclei). The fact that
 $e_\alpha$ is much smaller than $e_s^{(n)}+e_s^{(p)}$ indicates
 the very large energy gain of the $\alpha$-like quartet.
 Moreover, while the $\alpha$-like quartet interaction is attractive,
 the quasi-pair interaction is repulsive.
 The characteristic patterns in Figs. \ref{fig6} and \ref{fig7} are
 due to the repulsive interaction between the quasi-pairs
 (a quasi-pair transfers isospin 1). These are contrast with
 the inconspicuous effect of the multi-quartet structure on the energy
 $E(m \alpha)$, shown in Fig.~\ref{fig1}.
  (The $\alpha$-like quartet transfers no quantum number
 other than the nucleon number.)
 If there was no such great energy gain caused by the collective 
 $T=0$ $2n-2p$ correlations, the nuclear mass table would be different.

\section{Structure having an unpaired neutron and/or an unpaired proton}

   Before ending the second stage treating the multi-pair states,
 let us write our mass formula as
\begin{equation}
  B(A) = B_0(A)
       + E_{T=0}(A) + w_T(A) + w_v(A).        \label{eq:28}
\end{equation}
 We extend the $T$-dependent energy $W(A_0+m\alpha+2l:T=l)$ in Eq. (\ref{eq:24})
 to odd-$A$ nuclei and odd-odd nuclei, as we extended $E(A_0+m\alpha)$
 to $E_{T=0}(A)$, expressing it as 
\begin{equation}
  w_T(A) = a_{sym}(A) T^2 + b_{Wig}(A) T,   \label{eq:29}
\end{equation}
 where we permit $T$ to be a half integer for odd-$A$ nuclei.
 The last term, $w_v(A)$ in Eq. (\ref{eq:28}), represents the energy of
 unpaired nucleon(s). The subscript $v$ is the seniority quantum
 number.  It should be noted that in the mass formula (\ref{eq:28}),
 the energy $w_v(A)$ of an odd-$A$ (or odd-odd) nucleus is measured
 from the base {\it curve} given by (\ref{eq:29}).

\subsection{Shifted quasi-particle energy for odd-mass nuclei}

   The strength of the pairing correlations in an odd-$A$ nucleus is
 usually evaluated with the odd-even mass difference.  We define it
 using $W(A)$ of Eq. (\ref{eq:12}) in the same form as Eq. (\ref{eq:14}),
\begin{equation}
  \Delta(A=A_0+m\alpha+2l+1)
  = W(A)  - ( W(A-1) + W(A+1) ) /2. \label{eq:30}
\end{equation}
 This relation is illustrated in Fig. \ref{fig2}.
 Substituting the relation (\ref{eq:29}) for $W(A-1)$ and $W(A+1)$,
 we obtain the approximate relation
\begin{equation}
   \Delta(A=A_0+m\alpha+2l+1)
   \approx W(A) - ( w_{T=l+1/2}(A) + a_{sym}(A)/4 ). \label{eq:31}
\end{equation}
 The energy $w_{v=1}(A)$ for an odd-$A$ nucleus with $A=A_0+m\alpha+2l+1$
 in Eq. (\ref{eq:28}) is given by
\begin{eqnarray}
  w_{v=1}(A) & \equiv & W(A) - w_{T=l+1/2}(A)     \nonumber \\
   & \approx & \Delta(A) + \frac{1}{4} a_{sym}(A).   \label{eq:32}
\end{eqnarray}
 The energy shift $a_{sym}(A)/4$ is inevitable when we measure the energy
 $w_{v=1}(A)$ from the base curve (\ref{eq:29}).
 
\begin{figure}[b]
\begin{center}
\includegraphics[width=6.8cm,height=7.2cm]{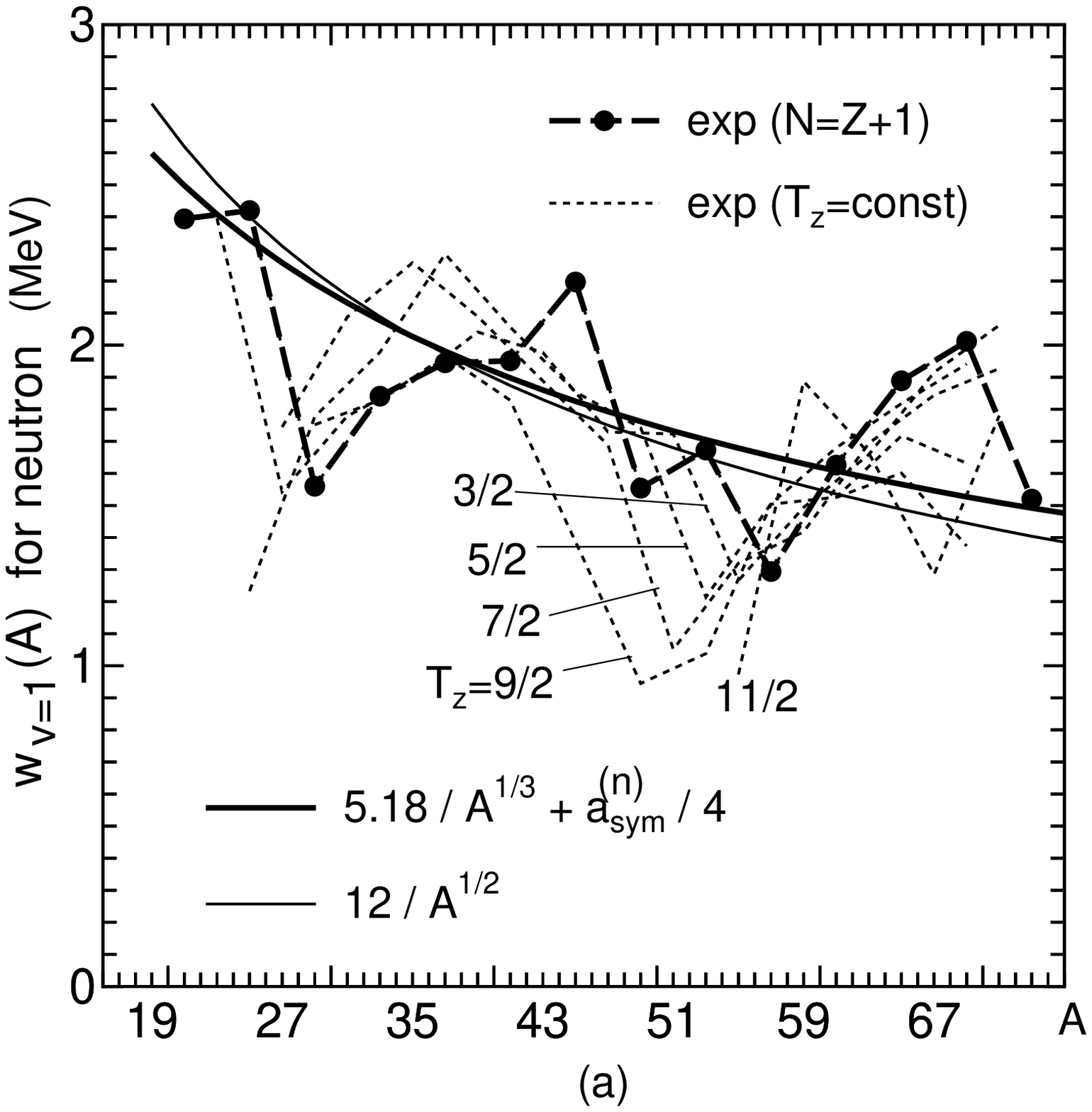}
\includegraphics[width=6.8cm,height=7.2cm]{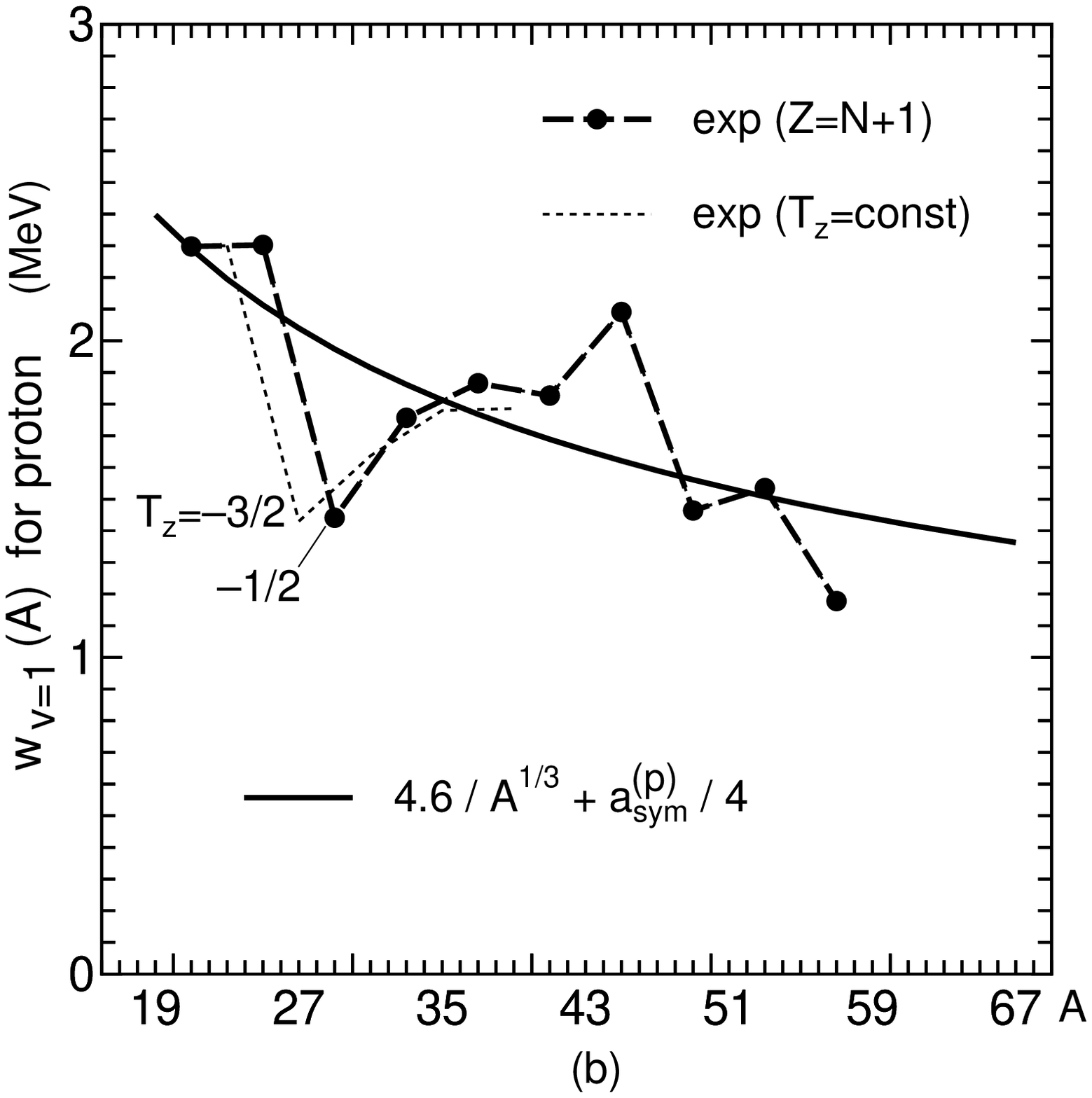}
  \caption{Energies $w_{v=1}(A)=d_n(A)$ for odd-$N$ nuclei and
           $w_{v=1}(A)=d_p(A)$ for odd-$Z$ nuclei.}
  \label{fig8}
\end{center}
\end{figure}

  In the last stage, we consider $A_0+ m\alpha +2l+1$ nuclei
 with $T=l+1/2$, which have the structure
\begin{equation}
 |A=A_0+m\alpha+2l+1 \rangle \propto c^\dagger (S^\dagger)^l
          |\Phi_0(A_0+m\alpha) \rangle . \label{eq:33}
\end{equation}
 This structure is expressed approximately as a direct product of
 the three modules $|\Phi_0(A_0+m\alpha) \rangle$, $(S^\dagger)^l$, and
 the last odd nucleon $c^\dagger$.
 We regard the multi-pair structure $(S^\dagger)^l$ as the pairing
 superfluid structure as usual.  After the Bogoliubov transformation,
 the odd-$A$ nucleus is regarded as the one quasi-particle state,
\begin{equation}
 |A=A_0+m\alpha+2l+1 \rangle = a^\dagger
          |0(lS)\otimes 0(A_0+m\alpha) \rangle . \label{eq:34}
\end{equation}
 In this picture, the energy $w_{T=l}(A)$ given in Eq. (\ref{eq:28}) represents
 the energy of the pairing superfluid state $|0(lS) \rangle$,
  which is the vacuum for the quasi-particle $a^\dagger$,
 and the quantity $\Delta (A)$ in Eq. (\ref{eq:30})
 can be regarded as the quasi-particle energy.
  Let us rewrite the ``shifted quasi-particle energy" $w_{v=1}(A)$ as
\begin{equation}
  d_n(A) = \Delta_n(A) + a_{sym}^{(n)}/4, \quad
  d_p(A) = \Delta_p(A) + a_{sym}^{(p)}/4. \label{eq:35}
\end{equation}
 Experimental values of $w_{v=1}(A)$ calculated with the experimental
 values of $E_{T=0}(A)$ in Eq. (\ref{eq:28}) are plotted
 in Fig. \ref{fig8}.
  With the approximate relation (\ref{eq:32}), we can parameterize
 the quantity $\Delta(A)$ in the same form ($\propto A^{-1/3}$)
 as that of the microscopic mass formula \cite{Duflo}, {\it i.e.},
\begin{equation}
  \Delta_n(A) = 5.18/A^{1/3}, \quad 
  \Delta_p(A) = 4.6/A^{1/3}.   \label{eq:36}
\end{equation}
 The neutron value, $\Delta_n(A)$, is equal to that given in Ref. 4),
 and the proton value, $\Delta_p(A)$, is smaller than $\Delta_n(A)$.
 It should be noted that $\Delta(A)$ is a measure of the $T=1$ pair
 correlations and $W(A:T=1)=e_s$ is approximately a measure of
 the $T=0$ $n-p$ correlations between the $T=1$ pairs \cite{Kaneko,Janecke}.
 This leads to the different $A$ dependences of $\Delta(A)$ and $e_s$.
 
  According to Ref. 39), because the symmetry energy contribution
 is cancelled by the curvature contribution from a smooth density of states
 in the Strutinsky method, the three-point odd-even mass difference
 $\Delta_n(A)$ in Eq. (\ref{eq:30}) is a good indicator of the pairing gap,
 which is approximately equal to the quasi-particle energy. 
 It is notable that, in contrast to the $A$-dependence $5.18/A^{1/3}$ of
 $\Delta_n(A)$, the $A$-dependence of $w^{(n)}_{v=1}(A)=d_n(A)$ can
 be expressed as $12/\sqrt{A}$, as shown in Fig. \ref{fig8}(a).
 The curve $12/\sqrt{A}$ is known to represent the $A$-dependence of the pairing
 energy of the semi-empirical mass formula  \cite{Zeldes}, which is estimated
 with the four-point odd-even mass difference.
  The shifted quasi-particle energy $d_n(A)$, therefore, corresponds to
 the pairing energy of the semi-empirical mass formula or the four-point
 odd-even mass difference.
 The classical mass formulas, having the pairing energy term $\delta_{pair}$
 and the symmetry energy term $a_T T^2$, lead to the relation
 $\Delta_n(A) \approx \delta_{pair}- a_T/4$. Combining this relation and
 Eq. (\ref{eq:35}), we confirm the equivalence $d_n=\delta_{pair}$.
  Equation (\ref{eq:35}) indicates that the so-called pairing energy
 $d_n=\delta_{pair}$ contains a symmetry energy contribution. 
 We can now distinguish the two curves $5.18/A^{1/3}$ and $12/\sqrt{A}$:
 The former represents the three-point odd-even mass difference  $\Delta_n(A)$
 (which is the quasi-particle energy or the pairing gap), and the latter
 represents the four-point odd-even mass difference equal to $d_n(A)$,
 including the symmetry energy contribution $a_{sym}^{(n)}/4$.
 
\subsection{Seniority $v=2$ states of odd-odd nuclei}

\begin{figure}[t]
\begin{center}
\includegraphics[width=7cm,height=7cm]{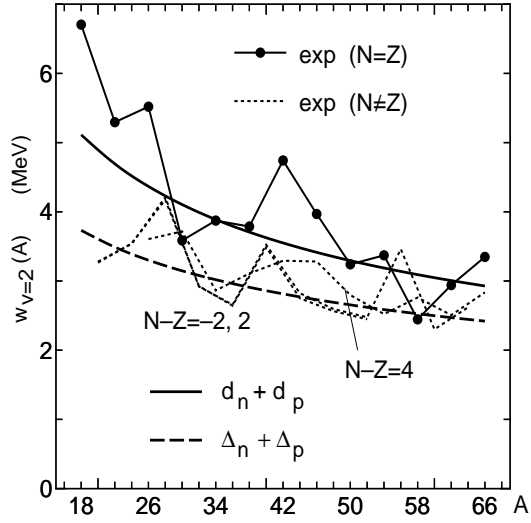}
  \caption{Energies $w_{v=2}(A)$ for odd-odd nuclei.}
  \label{fig9}
\end{center}
\end{figure}

   The remaining task is to determine whether the mass formula (\ref{eq:28})
 is effective for odd-odd nuclei.  The ground state of an odd-odd
 nucleus is the seniority $v=2$ state, except in the case of some $N=Z$ nuclei.
 (The exceptional state with $v=0$ and $T=1$ is the $1n1p$ pair state
 $S_{K=0}^\dagger |\Phi_0(A_0+m\alpha) \rangle$ considered in Fig. \ref{fig5}.)
 The seniority $v=2$ state is composed of a quasi-neutron and a quasi-proton,
\begin{equation}
 |A=A_0+m\alpha+2l+n+p \rangle = a_n^\dagger a_p^\dagger
          |0(lS)\otimes 0(A_0+m\alpha) \rangle . \label{eq:37}
\end{equation}
 The energy $w_{v=2}(A)$ for this state is defined
 by $w_{v=2}(A)=W(A)-w_{T=l}(A)$.
 Let us evaluate its experimental value in a manner similar to Eq. (\ref{eq:30}):
\begin{eqnarray}
 & {} & w_{v=2}(A=A_0+m\alpha+2l+n+p)    \nonumber \\
 & {} & \ \ \ = W(N,Z) - ( W(N-1,Z-1) + W(N+1,Z+1) ) /2. \ \   \label{eq:38}
\end{eqnarray}
 The calculated values are plotted in Fig. \ref{fig9}.
 It is seen that there is a difference between the odd-odd $N=Z$ nuclei
  and the other odd-odd nuclei.  The data indicate the relations
\begin{eqnarray}
        w_{v=2}(N=Z) \approx d_n + d_p ,  \label{eq:39} \\
  w_{v=2}(N \neq Z) \approx \Delta_n + \Delta_p .   \label{eq:40}
\end{eqnarray}

 The parameters $\Delta_n$ and $\Delta_p$ in Eq. (\ref{eq:36})
 [$d_n$ and $d_p$ in Eq.~(\ref{eq:35})]
 fitted for the odd-$A$ nuclei can reproduce the experimental energies
 $w_{v=2}$ of odd-odd nuclei, though it is not clear why $w_{v=2}(N=Z)$
 is different from $w_{v=2}(N \neq Z)$.
 This point is possibly related to the condition that there is
 no Cooper pair in odd-odd $N=Z$ nuclei, while the other odd-odd nuclei
 have one or more Cooper pairs.
  Sometimes, correction terms are added to mass formulas for odd-odd nuclei.
 The correction for the odd-odd $N=Z$ nuclei is included
 in Eq. (\ref{eq:39}) in contrast to Eq. (\ref{eq:40})
 for odd-odd $N \neq Z$ nuclei.  We ignore another correction,
 which represents an additional $n-p$ interaction,
 because the deviations from the fitted curves in Fig. \ref{fig9} are
 of a similar or smaller magnitude than the deviations
 in Figs. \ref{fig5}--\ref{fig9}.

\section{Concluding remarks}

   We have shown the essential role of the  $T=0$ $2n-2p$ correlations
 in the nuclear mass by considering concrete nuclear structure based on
 the $jj$ coupling shell model. 
 We find that explicitly taking account of the effects
  of the $T=0$ $2n-2p$ correlations,
 which have been overlooked in the past,
  is important for understanding
 the nuclear mass formula.  We have rearranged the mass formula
 by treating the $T=0$ $2n-2p$ correlations and the $T=1$ pair correlations
 as the most important correlations in nuclei.  Let us write it again:
\begin{equation}
  B(A) = B_{VSC}(A) + \delta U_{pot}(A)
       + E_{T=0}(A) + w_T(A) + w_v(A).  \nonumber
\end{equation}
 We have discussed the fact that the last three terms $E_{T=0}(A)$, $w_T(A)$ and
 $w_v(A)$ represent the three modules of the structured wave functions
 sketched in Eqs. (\ref{eq:10}), (\ref{eq:22}) and (\ref{eq:33}).
 The systematic formulation of the $T=0$ $2n-2p$ and $T=1$ pair correlations
 on the same footing makes it clear that the energy $E_{T=0}(A)$ of
 the multi-quartet structure should be added to the energy $w_T(A)$
 of the multi-pair structure.  The $T=0$ energy plane $E_{T=0}(A)$ supplies
 the base level for the measurement of the $T$-dependent energy $w_T(A)$.
 The interacting boson model for the $T=1$ Cooper pair on the $\alpha$-like
 superfluid base provides a structural explanation for the origins
 of the symmetry energy and Wigner energy.
 The two standard curves $5.18/A^{1/3}$ and $12/\sqrt{A}$ for the pairing
 energy are distinguished and identified as representing the quasi-particle
 energy or the pairing gap (three-point odd-even mass difference) and
 the shifted quasi-particle energy
 (four-point odd-even mass difference), respectively.

   The $E_{T=0}(A)$ term as the base level affects the binding energies
 of all nuclei.  Adding $E_{T=0}(A)$ to existing mass formulas
 could improve the precision.  We can estimate the precision
 using the parameters in Eqs. (\ref{eq:26}), (\ref{eq:27}) and (\ref{eq:36})
 and the experimental values of $E_{T=0}(A)$.
  The average of the root-mean-square (rms) errors estimated
 is 1.42 MeV for even-even nuclei, 1.37 MeV for odd-$A$ nuclei, and
 1.11 MeV for odd-odd nuclei.  These values are, of course, larger than
 the rms errors for modern mass formulas. The average of the rms errors
 for the FRDM, for instance, is 1.08 MeV for even-even nuclei,
 1.13 MeV for odd-$A$ nuclei, and 1.12 MeV for odd-odd nuclei
 in the region $17 \le Z,N \le 36$.
 However, it should be noted that these FRDM values are larger than
 the average of the rms errors for all nuclei, 0.67 MeV.
 This suggests a flaw in the FRDM mass formula for $N \approx Z$ nuclei.
 The advantage of our treatment is clear if we consider nuclei
 near the $N=Z$ line. 
 For $T<4$ nuclei, the average of the rms errors becomes 0.63 MeV
 for even-even nuclei and 1.07 MeV for odd-$A$ nuclei.  The good
 parallelism from the $T=0$ line to the $T=3$ line in Fig. \ref{fig4} reveals
 this mechanism.  By contrast, the FRDM mass formula does not
 show such a reduction when the number of $T=|N-Z|/2$ is limited.
 There seems to be room to take into account the energy $E_{T=0}(A)$
 of the fundamental $T=0$ $2n-2p$ correlated structure
 in the modern mass formulas.



%

\end{document}